\documentclass[journal]{IEEEtran}

\ifCLASSINFOpdf
  \usepackage[pdftex]{graphicx}
  \graphicspath{{./Figures/}{./photos/}}
  \DeclareGraphicsExtensions{.pdf,.png,.jpg}
\else
  \usepackage[dvips]{graphicx}
  \graphicspath{{./Figures/}{./photos/}}
  \DeclareGraphicsExtensions{.pdf,.png}
\fi

\usepackage{amsmath}
\ifCLASSOPTIONcompsoc
 \usepackage[caption=false,font=normalsize,labelfont=sf,textfont=sf]{subfig}
\else
 \usepackage[caption=false,font=footnotesize]{subfig}
\fi
\usepackage{dblfloatfix}
\usepackage{url}

\usepackage{amsfonts}

\usepackage{flushend}

\usepackage{enumitem}

\newcommand{\sgn}{\textrm{sgn}}
\newcommand\given[1][]{\:#1\vert\:}

\usepackage{algorithm}
\usepackage{algorithmic}
\usepackage{siunitx}

\makeatletter
\newcommand\fs@norules{\def\@fs@cfont{\bfseries}\let\@fs@capt\floatc@ruled
  \def\@fs@pre{}%
  \def\@fs@post{}%
  \def\@fs@mid{\kern3pt}%
  \let\@fs@iftopcapt\iftrue}
\makeatother
\floatstyle{norules}
\restylefloat{algorithm}

\begin{document}
%
\title{High-Order Signed Distance Transform of Sampled Signals}
%
%
%


\author{
  Bryce~A.~Besler, 
  Tannis~D.~Kemp,
  Nils~D.~Forkert, 
  Steven~K.~Boyd

\thanks{This work was supported by the Natural Sciences and Engineering Research Council (NSERC) of Canada, grant RGPIN-2019-4135.}
\thanks{B.A. Besler and T. D. Kemp are in the McCaig Institute for Bone and Joint Health, University of Calgary Canada. N.D. Forkert is with the Department of Radiology and the Hotchkiss Brain Institute, University of Calgary, Canada. S.K. Boyd is with the Department of Radiology and McCaig Institute for Bone and Joint Health, University of Calgary Canada e-mail: skboyd@ucalgary.ca}

}

%
%

\markboth{}%
{Besler \MakeLowercase{\textit{et al.}}: High-Order Signed Distance Transform of Sampled Signals}
%



\maketitle

\begin{abstract}
Signed distance transforms of sampled signals can be constructed better than the traditional ``exact'' signed distance transform.
Such a transform is termed the high-order signed distance transform and is defined as satisfying three conditions: the Eikonal equation, recovery by a Heaviside function, and has an order of accuracy greater than unity away from the medial axis.
Such a transform is an improvement to the classic notion of an ``exact'' signed distance transform because it does not exhibit artifacts of quantization.
A large constant, linear time complexity high-order signed distance transform for arbitrary dimensionality sampled signals is developed based on the high order fast sweeping method.
The transform is initialized with an exact signed distance transform and quantization corrected through an upwind solver for the boundary value Eikonal equation.
The proposed method cannot attain arbitrary order of accuracy and is limited by the initialization method and non-uniqueness of the problem.
However, meshed surfaces are visually smoother and do not exhibit artifacts of quantization in local mean and Gaussian curvature.
\end{abstract}


%
\IEEEpeerreviewmaketitle

\section{Introduction}
\label{sdt:sec:introduction}
It was recently demonstrated that distance transforms of sampled signals produce a quantized distance map~\cite{besler2020artifacts}.
This originates from the sampling period reflecting through the transform, discretizing the distance metric.
The quantization is independent of sampling period and cannot be corrected by increasing the resolution of the image.
Gradients of the distance map exhibit banding since neighbors in the finite difference operator are spaced at multiples of the sampling period.
As a result, algorithms that depend on computing gradients in the signed distance transform can be very noisy.

While the signed distance transform has many uses, the applications most affected by quantization are surface evolution problems initialized from image data.
A signed distance map is an implicit representation of a surface convenient for computing geometric flows as it allows the topology of the object to change without explicit splitting and merging rules~\cite{osher1988fronts}.
The signed distance transform of a binary image is sometimes used to initialize the surface for computing flows~\cite{besler2018bone}.
However, if the initialization of the surface is not at least second-order accurate, the error in mean curvature is independent of sample spacing~\cite{coquerelle2016fourth}.
This paper is concerned with developing a high-order signed distance transform of sampled signals.

\section{Review of Literature}
\label{sdt:sec:literature}

The problem is to construct a signed distance map given a binary image.
There are two definitions used for the signed distance transform, the first defines the embedding on the surface to be zero:
\begin{equation}
  \label{sdt:eqn:sdt_bad}
  SDT(x, I) = \begin{cases}
    +d(x, I^c) & \text{ if } x \in \Omega^- \\
    -d(x, I) & \text{ if } x \in \Omega^+ \\
    0 & \text{ if } x \in \Gamma
 \end{cases}
\end{equation}
where $\Omega^+$ is the set of foreground voxels, $\Omega^-$ is the set of background voxels, $I^c$ is the complement of the image, and $\Gamma$ is the set of voxels on the surface with one neighbor of an opposite label.
Alternatively, the zero condition on the surface can be removed:
\begin{equation}
  \label{sdt:eqn:sdt}
  SDT(x, I) = \begin{cases}
    +d(x, I^c) & \text{ if } x \in \Omega^- \\
    -d(x, I) & \text{ if } x \in \Omega^+
  \end{cases}
\end{equation}
The main distinction is whether a set of voxels on the surface are assigned exactly zero or if the surface is implicitly located between the edge of the surface voxels.
The implicit definition of the surface given by Equation~\ref{sdt:eqn:sdt} will be used for this work as it is desirable for curve evolution problems.

In many cases, the signed distance transform is generated by combining the distance transform of the foreground and background, so the literature review will treat the distance transform and signed distance transform as essentially interchangeable.

\subsection{Sliding Kernel Methods}
Originally, distance transforms were computed by sliding a kernel over the image that approximates a distance metric.
Distances between neighbors could be estimated using the kernel, allowing the computation of global distance by propagating local distances.
Kernel methods originated with Rosenfeld~\cite{rosenfeld1966sequential,rosenfeld1968distance} and neighborhoods were designed that were provably close to the Euclidean distance~\cite{danielsson1980euclidean,borgefors1986distance}.
Further improvements make the kernel separable~\cite{ragnemalm1993euclidean}.
These techniques were always approximations, leading to the desire to compute exact distances efficiently.
This gives rise to the exact distance transform, meaning the distance metric is the Euclidean norm measured exactly from one sample to another.

\subsection{Voronoi Diagrams}
Efficient methods based on the Voronoi diagram~\cite{voronoi1908nouvelles} emerged after the sliding kernel methods~\cite{breu1995linear,maurer2003linear}.
Voronoi cells extend away from foreground voxels on the surface, enclosing many background points.
Since Voronoi cells could be computed efficiently, the problem reduces to computing the intersection of background voxels with Voronoi cells, where the distance could be computed on the single element in the Voronoi cell.
While efficient, these methods also permitted the computation of exact Euclidean distances.

\subsection{Specialized Algorithms}
Many approaches have been developed for computing distance transforms beyond the two broad classes of sliding kernel methods and Voronoi diagrams.
Methods based on minimizing parabolas~\cite{saito1994new} and min-convolution~\cite{felzenszwalb2012distance} are particularly efficient.
Importantly, the artifact of quantization persists in all algorithms that compute an exact distance transform of a sampled grid.
Distance transforms better than the ``exact'' distance transform can be achieved in the sense that they have an increased order of accuracy.

Some attempts have been made to rectify this artifact.
Smooth distance transforms have been proposed where the minimum is replaced by a negative $\ell^p$-norm~\cite{brunet2016generalized}.
Alternatively, distances are computed from grey-scale images where gradient information already exists~\cite{kimmel1996sub,gustavson2011anti}.
These techniques make assumptions that do not hold for all applications, such as requiring a grey-scale image or modifying the zero level set of the distance transform.

\subsection{Model-Based Distance Transforms}
Finally, model-based distance transforms also exist.
In computer aided design, the distance transform of geometric shapes is known analytically and complex shapes can be formed by the intersection and union of these simple shapes~\cite{zhao2000implicit,museth2002level,baerentzen2005signed,pottmann2003geometry}.
Distance transforms of meshes~\cite{baerentzen2005robust,dapogny2012computation} could be applied by first meshing the binary image with marching cubes~\cite{lorensen1987marching}.
However, na\"ive implementation of the distance transform as the minimum over all triangles will also exhibit artifacts since the triangles are first-order structures.
Accurate normals are needed to define a smooth distance field, similar to the need for vertex normal interpolation when rendering meshes~\cite{gouraud1971continuous}.
These normals cannot be estimated accurately from binary images.
Additionally, going direct from binary image to signed distance map is preferable without having an intermediate meshing step.

\subsection{Eikonal Equation}
The Eikonal equation is found in many physical systems and generalizes the distance function. 
\begin{equation}
  \label{sdt:eqn:eikonal}
  | \nabla \phi | F  = 1
\end{equation}
It is a boundary value problem solving for the arrival time, $\phi$, of an initial front given a spatially varying speed function, $F$.
When $F=1$, this reduces to the distance transform.
The Fast Marching Method (FMM)~\cite{tsitsiklis1995efficient,sethian1996fast,chopp2001some,yatziv2006n}, Fast Sweeping Method (FSM)~\cite{tsai2003fast,zhao2005fast, zhao2007parallel}, and Fast Iterative Method (FIM)~\cite{jeong2008fast} are efficient algorithms for solving the Eikonal equation.

\subsection{Reinitialization Equation}
Sometimes called the unsteady Eikonal equation, the reinitialization equation is used to maintain a signed distance transform in computational fluid dynamics problems~\cite{sussman1994level,peng1999pde}.
\begin{equation}
  \phi_t + \sgn(\phi)\left(|\nabla \phi| - 1\right) = 0
\end{equation}
where $\sgn(\cdot)$ is the sign function.
This equation has been used for correcting the quantization artifact previously~\cite{besler2020artifacts} and computing a signed distance transform of grey-scale images~\cite{kimmel1996sub}.
However, the method is impractical, requiring hundreds or thousands of iterations over the entire image to converge~\cite{tsai2003fast,besler2020artifacts}.
The Eikonal equation will be used in this work as it is a technical means to improve the order of the signed distance transform and resolve the quantization artifact.

\section{Problem Definition}
\label{sdt:sec:definition}

A transform, $T$, is sought that computes the signed distance map, $\phi$, from a binary image, $I$.
\begin{equation}
  \label{sdt:eqn:transform}
  T(I) = \phi
\end{equation}
The n-dimensional domain of $\phi$ and $I$ is discrete, $\Omega \subset \mathbb{Z}^n$.
By convention, the inside of $\phi$ will be negative.

The computed distance map should satisfy the Eikonal Equation (Equation~\ref{sdt:eqn:eikonal}) with $F = 1$ and recover the original binary image.
Recovery of the original image can be stated precisely with the Heaviside function, $\theta$:
\begin{equation}
  \label{sdt:eqn:recovery}
  \theta( - \phi) = I
\end{equation}
The negative sign originates from the inside-is-negative convention.
Finally, the map should be sufficiently smooth away from shocks.
\begin{equation}
  \label{sdt:eqn:smoothness}
  \lVert \tilde{\phi} - \phi \rVert_p \leq C h^m
\end{equation}
where $\tilde{\phi}$ is the computed signed distance map, $\phi$ is the theoretical signed distance map, $m$ is the order of accuracy greater than 1, $h$ is the image spacing, $C$ is a constant independent of $h$, and $\lVert \cdot \rVert_p$ is an appropriate $\ell^p$ norm.
The solution can be no better than first-order accurate at shocks, but can be higher order accurate in smooth sections~\cite{zhao2005fast}.

\subsection{Uniqueness}
First, it is shown that $T$ does not permit a unique solution on a discrete domain given the conditions of Eikonal equation, Heaviside recovery, and smoothness.

\begin{figure}[t]
  \centering
  \begin{tabular}{cc}
    \subfloat[$I$]{
      \includegraphics[width=0.43\linewidth]{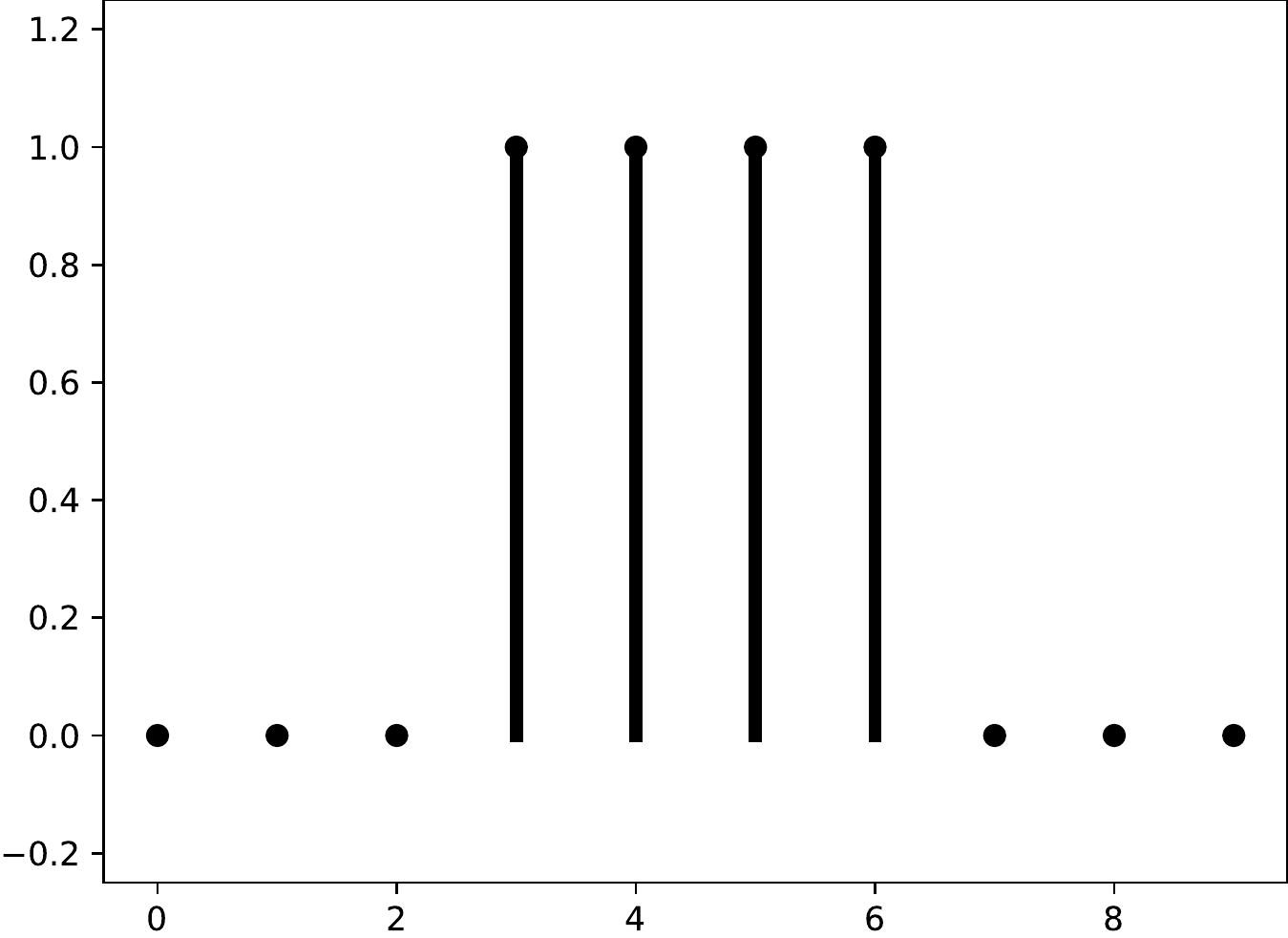}%
      \label{sdt:fig:unique:I}
    } & 
    \subfloat[$\phi$]{
      \includegraphics[width=0.43\linewidth]{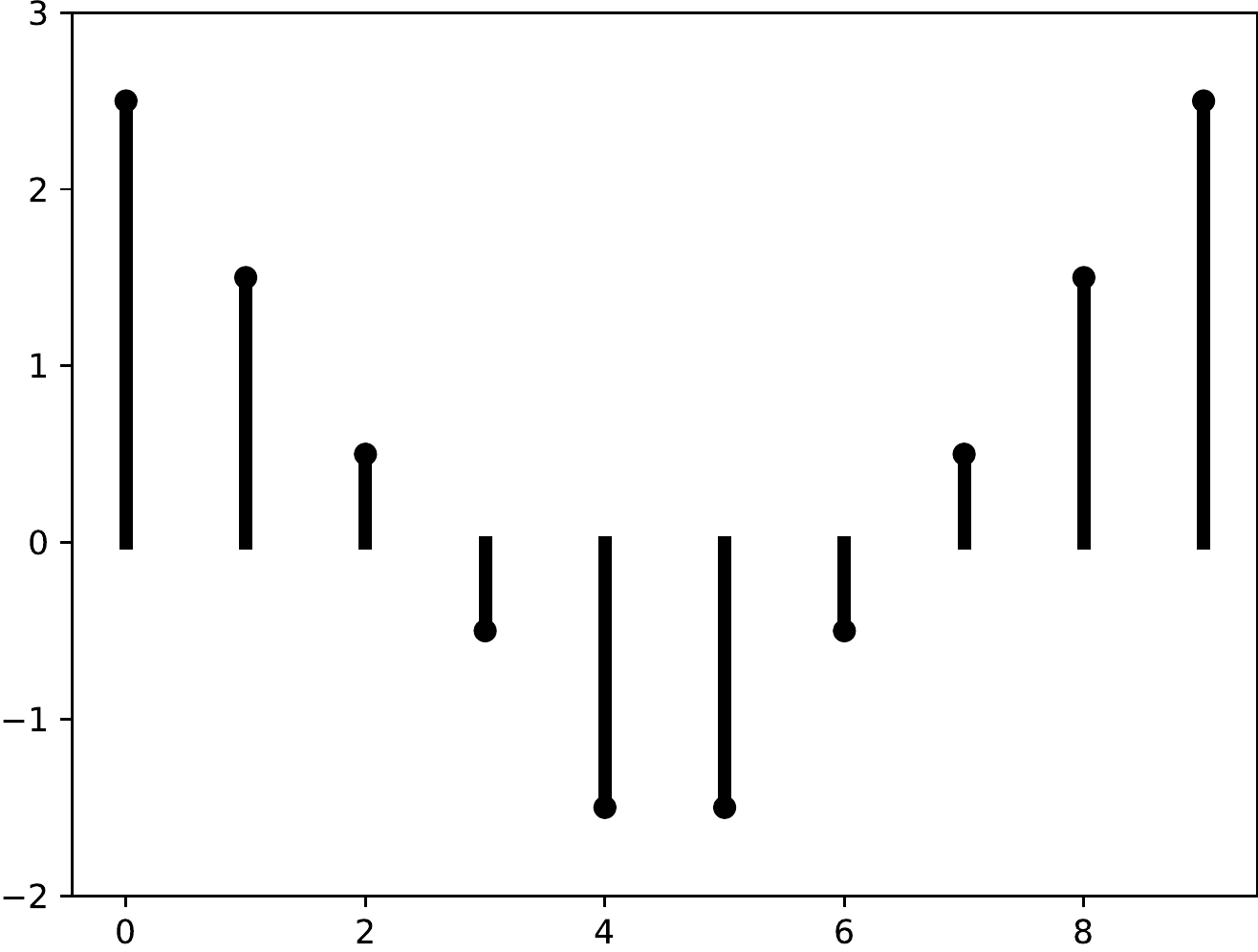}%
      \label{sdt:fig:unique:phi}
    } \\
    \subfloat[$\phi + 0.4$]{
      \includegraphics[width=0.43\linewidth]{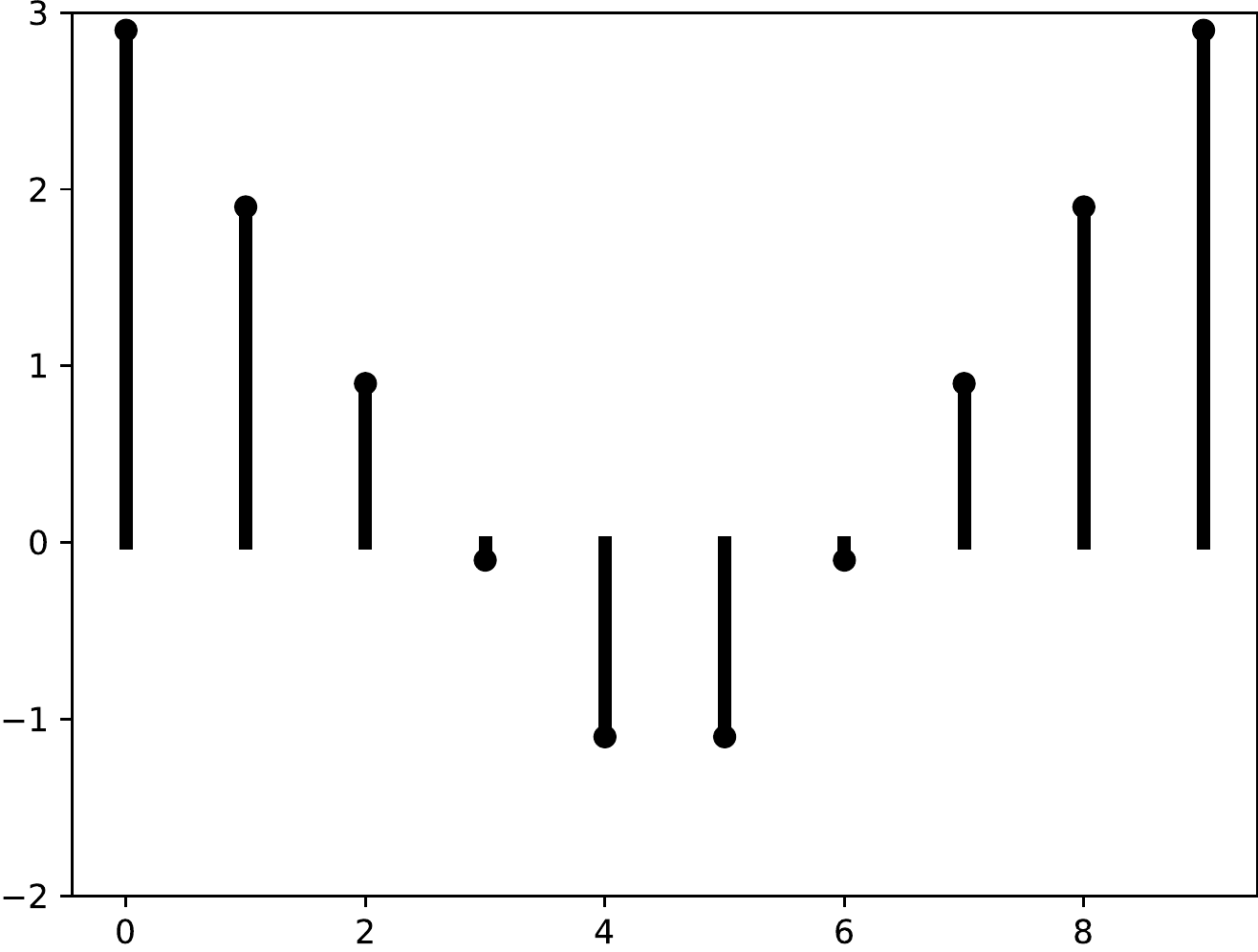}%
      \label{sdt:fig:unique:phi+}
    } & 
    \subfloat[$\phi - 0.4$]{
      \includegraphics[width=0.43\linewidth]{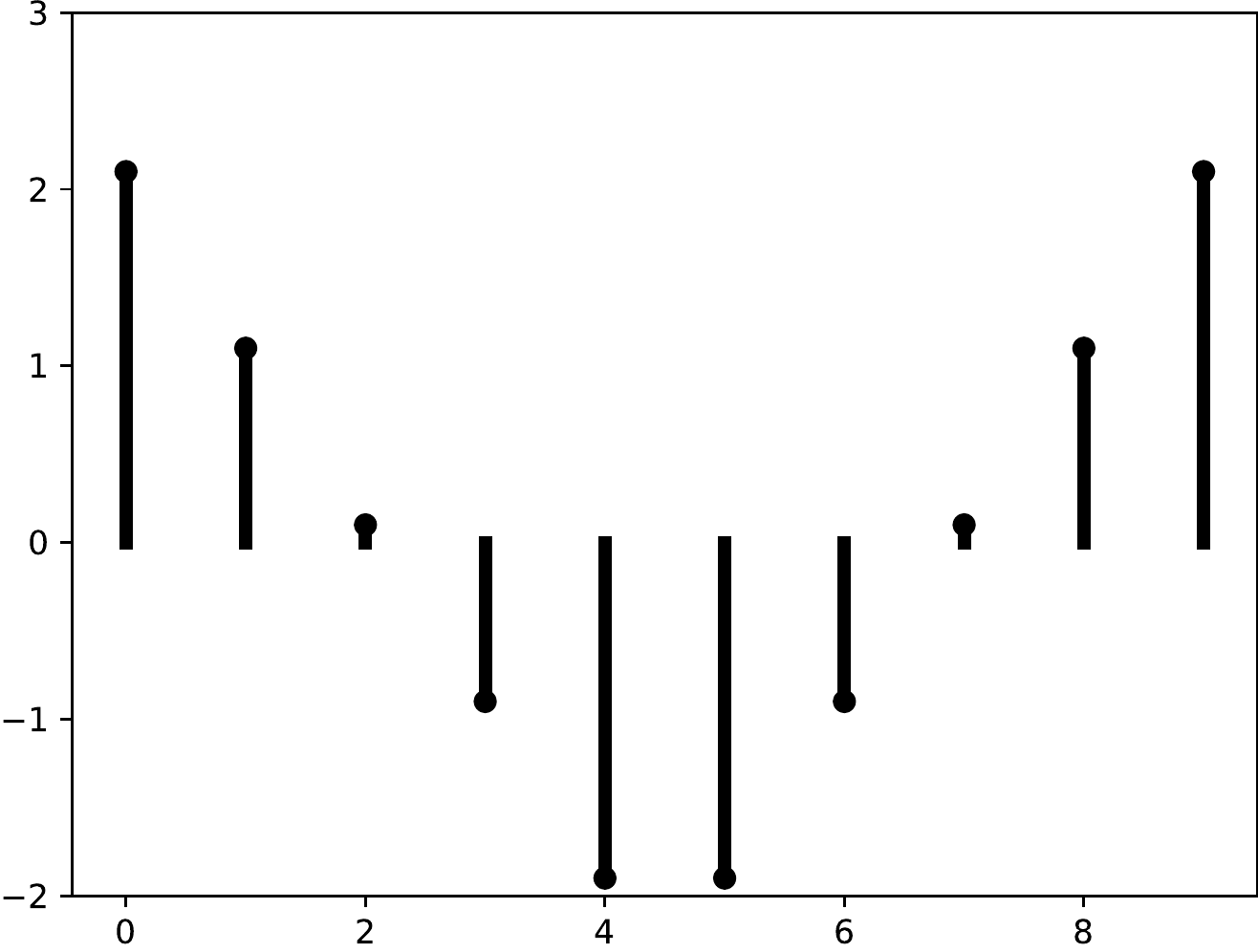}%
      \label{sdt:fig:unique:phi-}
    } \\
  \end{tabular}
  \caption{Multiple solutions exist given a single binary signal. Solutions \ref{sdt:fig:unique:phi}, \ref{sdt:fig:unique:phi+}, and \ref{sdt:fig:unique:phi-} are sub-voxel shifts all satisfying the Eikonal, Heaviside, and smoothness conditions.}
  \label{sdt:fig:unique}
\end{figure}

Consider a solution $\phi$ that is sufficiently smooth and satisfies the Eikonal equation sampled isotropically with sample period $h$.
Adding a constant to this solution does not modify the smoothness or Eikonal conditions.
A set of solutions can be generated that all satisfy the Heaviside condition.
\begin{equation}
  \mathcal{S} = \left\{ \phi + a \given |a| < h/2 \right\}
\end{equation}
$\mathcal{S}$ corresponds to the  set of all sub-voxel shifts in the embedding function.
This is visualized in Figure~\ref{sdt:fig:unique}.
The same argument holds in arbitrary dimensions by considering the n-sphere.
Ill-posedness is well understood in literature attempting to extract smooth surfaces from binary image data~\cite{whitaker2000reducing,lempitsky2010surface}.

\subsection{Causality}
\begin{figure}[b]
  \centering
    \includegraphics[width=0.5\linewidth]{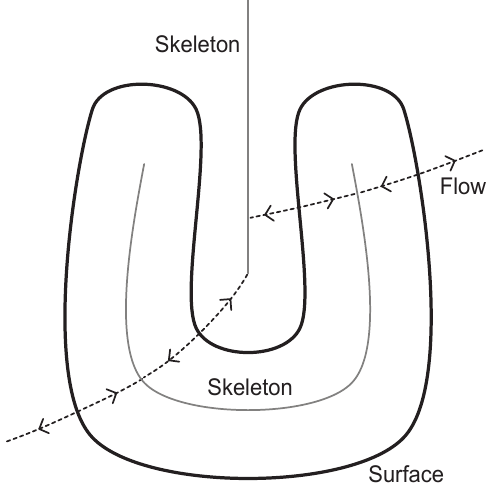}%
  \caption{Causality in a signed distance transform. The solution should flow from the surface outwards. A skeleton is drawn for inside and outside the curve. Figure adapted from~\cite{blum1967transformation}.}
  \label{sdt:fig:causal}
\end{figure}

Second, the solution should flow outward from the surface as illustrated in Figure~\ref{sdt:fig:causal}, consistent with the Grassfire transform of Blum~\cite{blum1967transformation}.

Already, reinitialization methods based on the unsteady Eikonal equation have information flow from the surface to the domain because advection is computed upwind~\cite{sussman1994level,peng1999pde,russo2000remark}.
The FSM, FMM, and FIM make use of causality to design efficient algorithms.
However, the initialization is critical.
If the narrowband around the surface is initialized with an exact distance transform, quantization will propagate through the analysis.

\subsection{Shocks at the Surface}
\label{sdt:subsec:suface_shocks}
Lastly, it is shown that binary images can be constructed that exhibit shocks along the surface.

\begin{figure}[t]
  \centering
  \begin{tabular}{cc}
    \subfloat[$I$]{
      \includegraphics[width=0.32\linewidth]{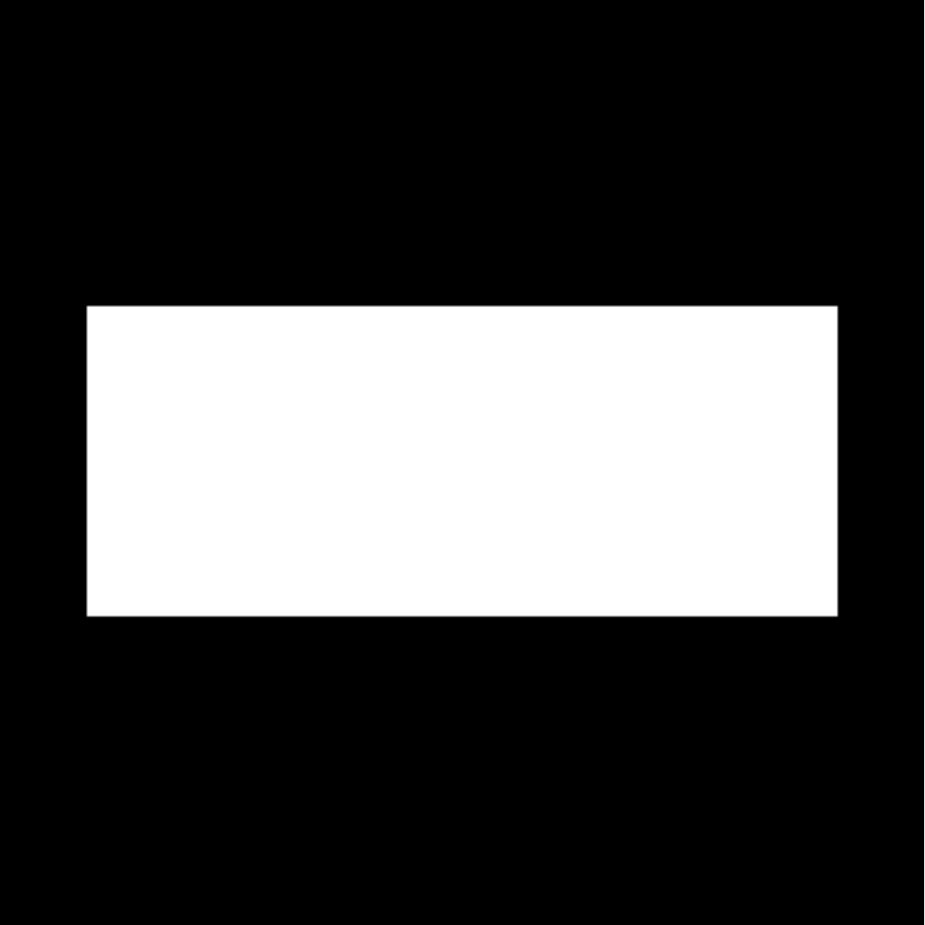}%
      \label{sdt:fig:shock:I}
    } & 
    \subfloat[$I$ Skeleton]{
      \includegraphics[width=0.32\linewidth]{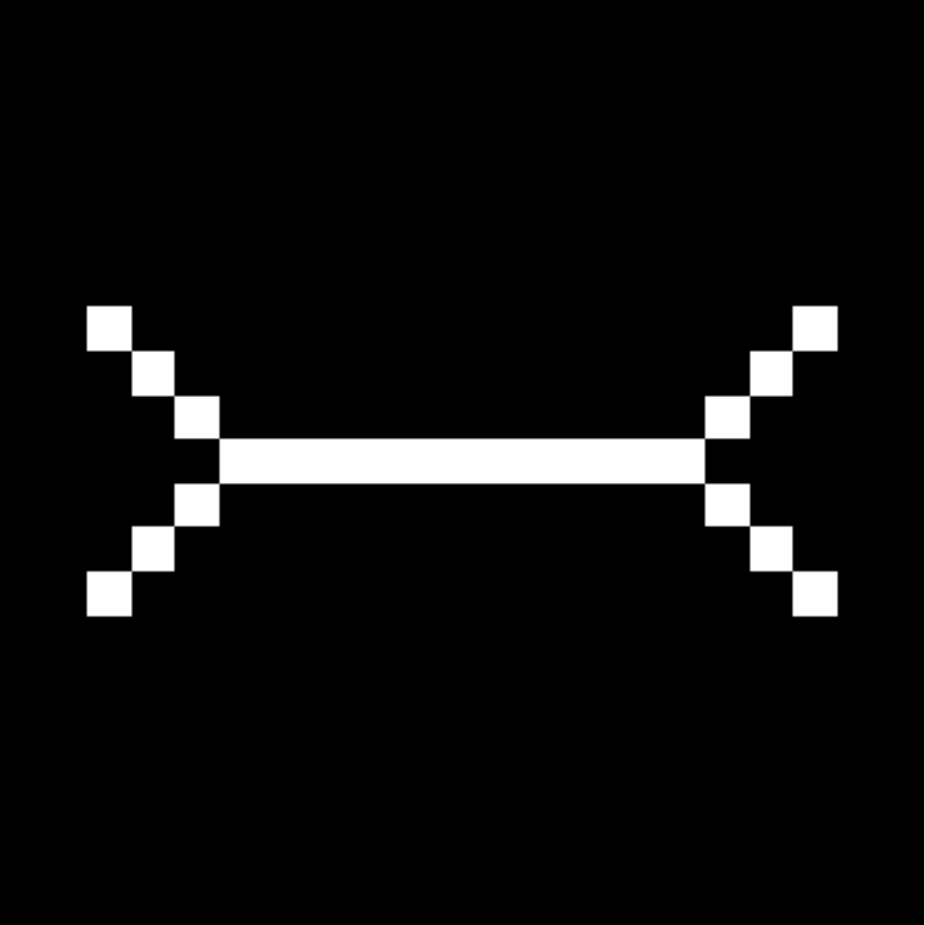}%
      \label{sdt:fig:shock:I_tf}
    } \\
    \subfloat[$J$]{
      \includegraphics[width=0.32\linewidth]{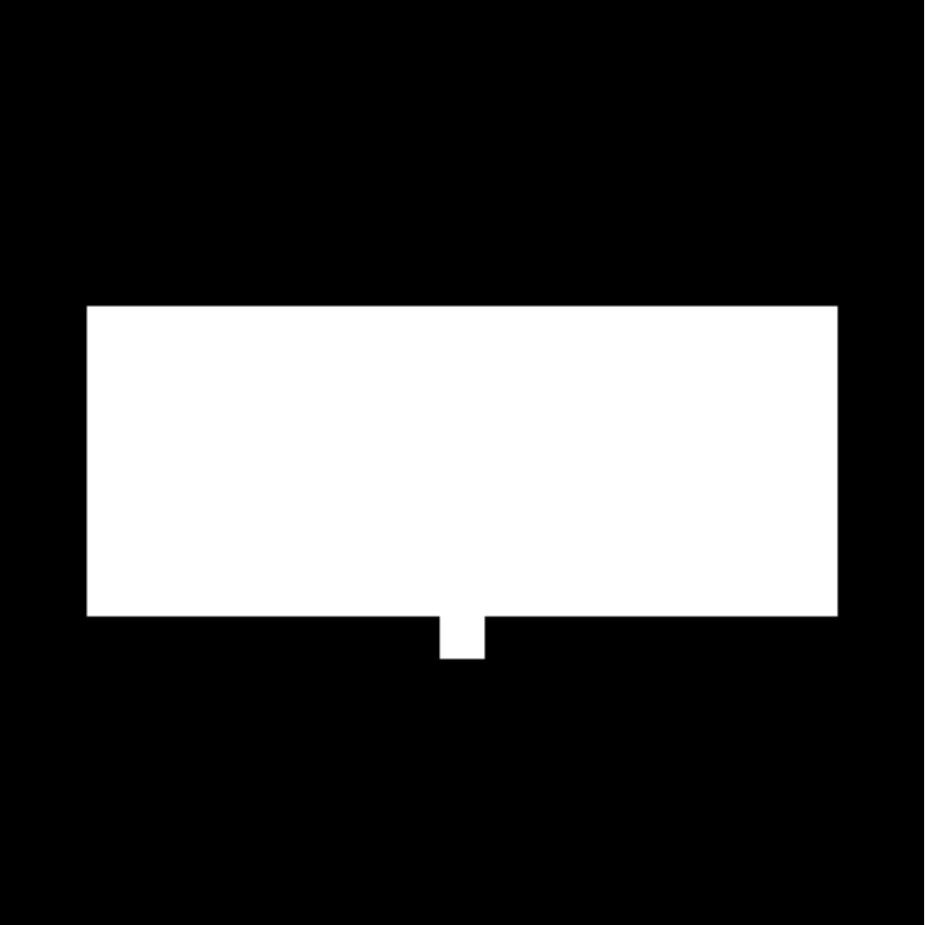}%
      \label{sdt:fig:shock:J}
    } & 
    \subfloat[$J$ Skeleton]{
      \includegraphics[width=0.32\linewidth]{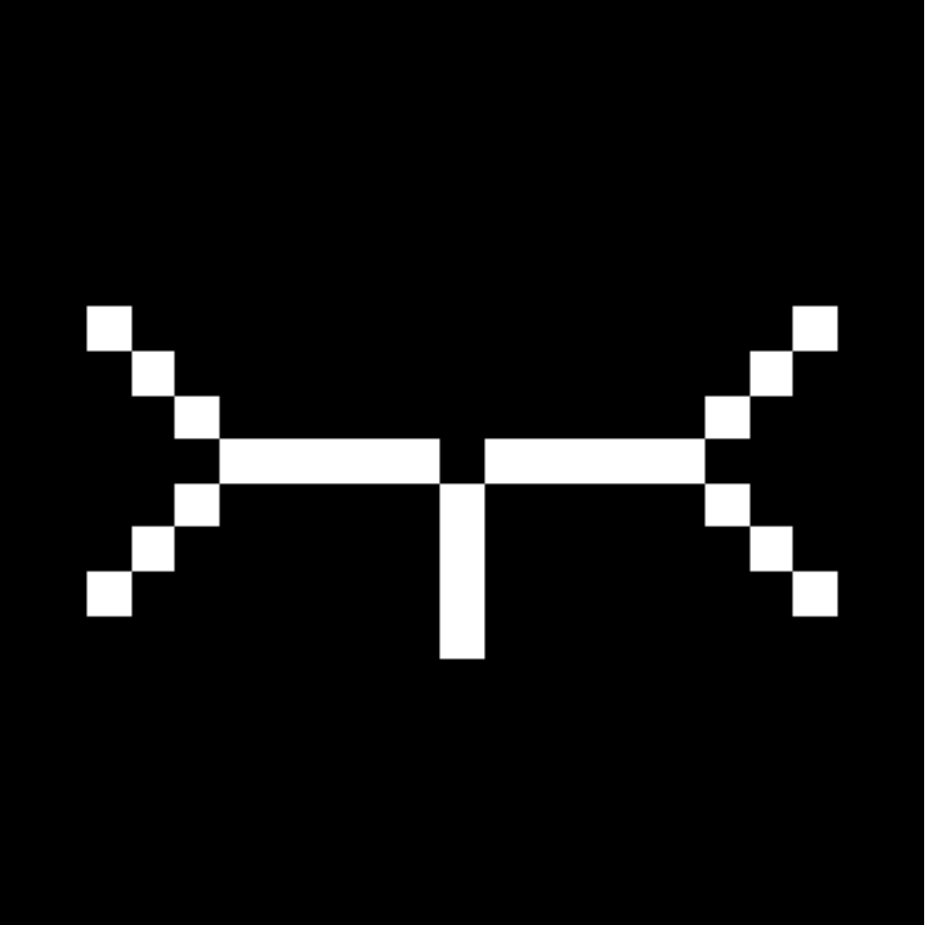}%
      \label{sdt:fig:shock:J_tf}
    } \\
  \end{tabular}
  \caption{Adding one voxel introduces a shock at the surface, connecting the new voxel to the original skeleton. Figure adapted from~\cite{august1999ligature}.}
  \label{sdt:fig:shock}
\end{figure}

The skeleton of a binary image corresponds to shocks in the signed distance transform~\cite{blum1967transformation,kimmel1995skeletonization,siddiqi2002hamilton} and can be analyzed through such tools.
It is well known that the skeleton is sensitive to the presence of voxels on the surface~\cite{august1999ligature,choi2004linear,chazal2004stability}.
This is demonstrated in Figure~\ref{sdt:fig:shock}. Since shocks can be no better than first-order accurate, propagating a solution away from a shock will limit the accuracy to first-order~\cite{zhang2006high}.


\section{High Order Signed Distance Transform}
\label{sdt:sec:transform}

\begin{figure*}
  \centering
  \begin{tabular}{cccc}
    \subfloat[Analytic]{
      \includegraphics[width=0.22\linewidth]{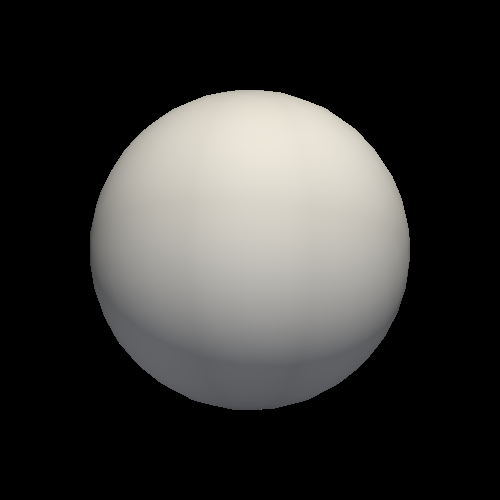}%
      \label{sdt:fig:result:analytic}
    } & 
    \subfloat[Binarized]{
      \includegraphics[width=0.22\linewidth]{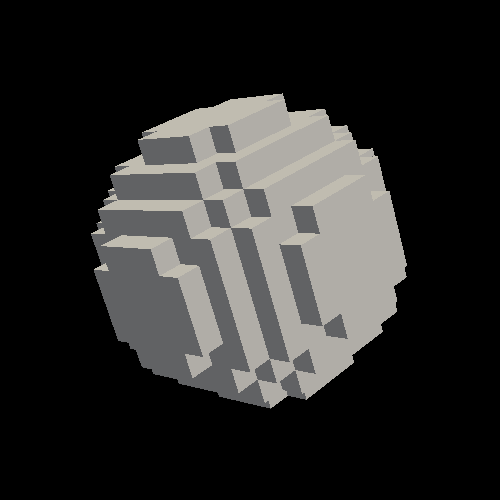}%
      \label{sdt:fig:result:binarized}
    } &
    \subfloat[Exact]{
      \includegraphics[width=0.22\linewidth]{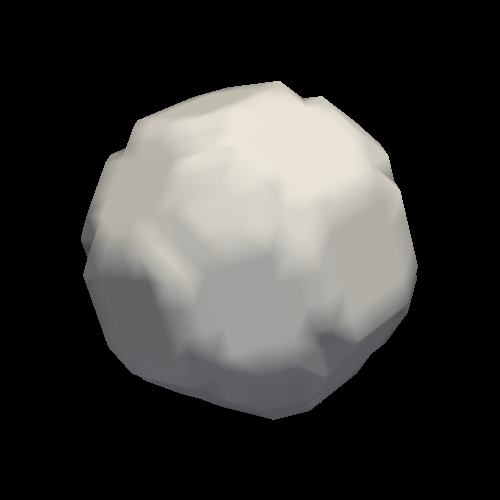}%
      \label{sdt:fig:result:standard}
    } & 
    \subfloat[High-Order]{
      \includegraphics[width=0.22\linewidth]{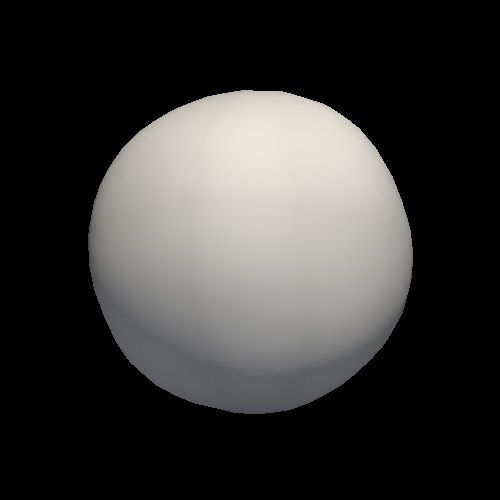}%
      \label{sdt:fig:result:proposed}
    } \\
    \subfloat[Analytic]{
      \includegraphics[width=0.22\linewidth]{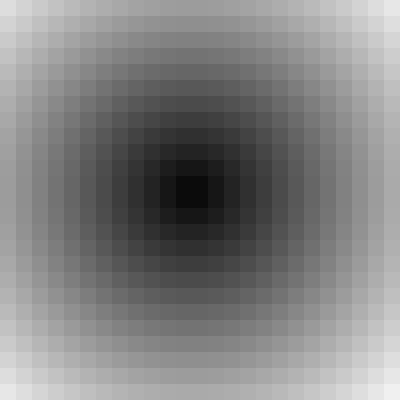}%
      \label{sdt:fig:result:analytic_slice}
    } & 
    \subfloat[Binarized]{
      \includegraphics[width=0.22\linewidth]{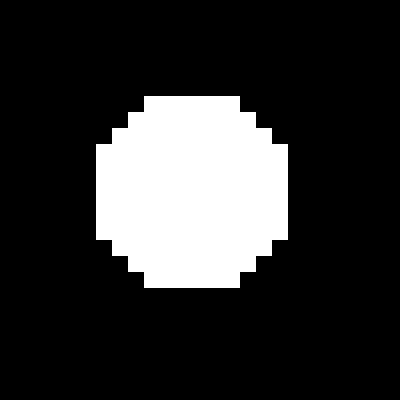}%
      \label{sdt:fig:result:binarized_slice}
    } &
    \subfloat[Exact]{
      \includegraphics[width=0.22\linewidth]{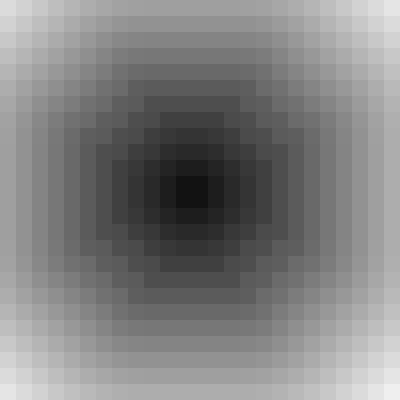}%
      \label{sdt:fig:result:standard_slice}
    } & 
    \subfloat[High-Order]{
      \includegraphics[width=0.22\linewidth]{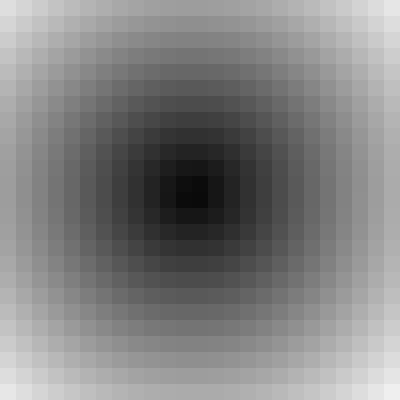}%
      \label{sdt:fig:result:proposed_slice}
    } 
  \end{tabular}
  \caption{Demonstration of the smoothness result. An analytic signed distance transform of a sphere ($r = 25$, $h = 4.0$) instantiated on a 25 x 25 x 25 grid. The transform is binarized (\ref{sdt:fig:result:binarized}) and the exact signed distance transform (\ref{sdt:fig:result:standard}) and proposed high order signed distance transform (\ref{sdt:fig:result:proposed}) computed. Zero level sets are visualized in the top row and central slices are visualized in the bottom row.}
  \label{sdt:fig:result}
\end{figure*}

An algorithm is presented for computing a high-order signed distance transform of a binary image.
The causality condition is enforced by sweeping the solution from the interface, outwards.
We term this technique the ``signed fast sweeping method'' based heavily on Zhao's fast sweeping method~\cite{tsai2003fast,zhao2005fast,zhang2006high}.
The results of the algorithm are shown in Figure~\ref{sdt:fig:result}.

\begin{figure*}[b]
  \centering
  \begin{tabular}{cccc}
    \subfloat[$I$]{
      \includegraphics[width=0.22\linewidth]{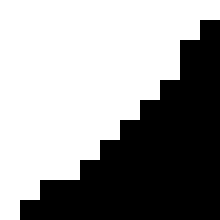}%
      \label{sdt:fig:init_sdt:I}
    } & 
    \subfloat[$SDT(I)$]{
      \includegraphics[width=0.22\linewidth]{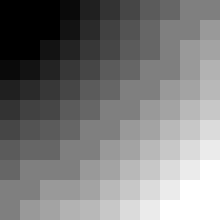}%
      \label{sdt:fig:init_sdt:forward}
    } &
    \subfloat[$SDT(I^c)$]{
      \includegraphics[width=0.22\linewidth]{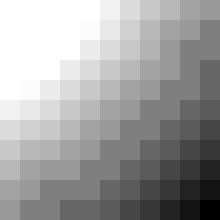}%
      \label{sdt:fig:init_sdt:backward}
    } & 
    \subfloat[$\phi_0$]{
      \includegraphics[width=0.22\linewidth]{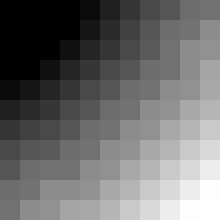}%
      \label{sdt:fig:init_sdt:average}
    }
  \end{tabular}
  \caption{Initialization of the signed distance map so the implicit surface goes through the edge of the voxels. In \ref{sdt:fig:init_sdt:forward} and \ref{sdt:fig:init_sdt:backward}, the grey band is exactly zero. By averaging these two transforms, a band of all zeros is avoided.}
  \label{sdt:fig:init_sdt}
\end{figure*}

\subsection{Initialization}
In the general solution of the Eikonal equation, the embedding is initialized from an analytic solution.
The distance transform of points or geometric objects are interpolated onto the narrowband and fixed.
However, the nature of the binary image does not permit a high order accuracy narrowband initialization.
Instead, the narrowband will be initialized with a standard signed distance transform and iteratively improved using the Eikonal equation.

The distance field is initialized with the Maurer exact signed distance transform~\cite{maurer2003linear}.
To avoid having a band of all zeros in the narrowband, the exact signed distance transform is computed twice, for the object and its complement, and the two transforms are averaged.
\begin{equation}
  \phi_0 = \frac{SDT(I) - SDT(I^c)}{2}
\end{equation}
The result is demonstrated in Figure~\ref{sdt:fig:init_sdt}.
This creates a signed distance map with no exactly zero voxels while having the implicit surface go through the voxel edges.
The problem now reduces to iteratively improving the embedding, keeping the sign the same.
This initialization still poses the quantization artifact, present in the narrowband of Figure~\ref{sdt:fig:init_sdt:average}.

\subsection{Solver}
The fast sweeping method is used to iteratively improve the signed distance transform~\cite{zhao2005fast}.
Since the FSM was originally designed for a positive-everywhere distance transform, it must be modified to handle the sign in the signed distance map.
This will be done by defining a ``signed minimum'' between two values.
Expressed verbosely, the signed minimum takes the minimum if the embedding is positive and the maximum if the embedding is negative.
That is, take the value that is closer to zero preserving the sign:
\begin{equation}
  \label{sdt:eqn:sgnmin}
  \text{sgnmin}(a, b, s) = s \cdot \min(s \cdot a, s \cdot b)
\end{equation}
where $\text{sgnmin}$ is the signed minimum function, $s$ represent the intended sign of the embedding, and $a$, $b$ are real numbers.
This will be used throughout the solver.

Typically, the sign of an embedding is computed by dividing through a regularized absolute value:
\begin{equation}
  s = \frac{\phi}{\sqrt{\phi^2 + \Delta x^2}}
\end{equation}
where $\Delta x$ is the image spacing.
However, the sign of each voxel can be computed quickly from the binary image:
\begin{equation}
  s = \begin{cases}
    +1 & x \in \Omega^- \\
    -1 & x \in \Omega^+
 \end{cases}
\end{equation}

\subsubsection{Numerical Approximation}
The Eikonal equation (\ref{sdt:eqn:eikonal}) is used to correct the initialization.
As is done in the FSM~\cite{zhao2005fast}, an upwind Godunov scheme is used to discretize the equation.
This is done in two dimensions for clarity, but extends to arbitrary dimensions.
The Godunov scheme is:
\begin{multline}
  \label{sdt:eqn:godunov}
  \begin{split}
    \max\left(\left[D^x_-\phi\right]^+, -\left[D^x_+\phi\right]^-\right)^2 + \\
    \max\left(\left[D^y_-\phi\right]^+, -\left[D^y_+\phi\right]^-\right)^2
   = \frac{1}{F^2}
  \end{split}
\end{multline}
where $D^i_\pm$ is the left or right sided first derivative along the i\textsuperscript{th} direction and the operators $\left(\cdot\right)^\pm$ are given by:
\begin{eqnarray}
  (x)^+ &=& \max(x, 0) \\
  (x)^- &=& \min(x, 0)
\end{eqnarray}
The upwind nature of the Godunov scheme enforces the causality condition.
If the embedding at the candidate voxel is positive, Equation~\ref{sdt:eqn:godunov} reduces to:
\begin{equation}
  \label{sdt:eqn:godunov_positive}
  \left[\left(\frac{\phi_{i,j} - \phi_{x,min}}{h_x}\right)^+ \right]^2 + \left[\left(\frac{\phi_{i,j} - \phi_{y,min}}{h_y}\right)^+ \right]^2 = \frac{1}{F^2}
\end{equation}
where:
\begin{eqnarray}
  \label{sdt:eqn:x_min}
  \phi_{x,min} &=& \min(\phi_{i+1,j}, \phi_{i-1,j}) \\
  \label{sdt:eqn:y_min}
  \phi_{y,min} &=& \min(\phi_{i,j+1}, \phi_{i,j-1})
\end{eqnarray}
However, if the embedding at the candidate voxel is negative, Equation~\ref{sdt:eqn:godunov} reduces to:
\begin{equation}
  \label{sdt:eqn:godunov_negative}
  \left[\left(\frac{\phi_{i,j} - \phi_{x,max}}{h_x}\right)^-\right]^2 + \left[\left(\frac{\phi_{i,j} - \phi_{y,max}}{h_y}\right)^-\right]^2 = \frac{1}{F^2}
\end{equation}
where:
\begin{eqnarray}
  \label{sdt:eqn:x_max}
  \phi_{x,max} &=& \max(\phi_{i+1,j}, \phi_{i-1,j}) \\
  \label{sdt:eqn:y_max}
  \phi_{y,max} &=& \max(\phi_{i,j+1}, \phi_{i,j-1})
\end{eqnarray}

\begin{algorithm}[t]
  \caption{Signed Upwind Quadratic Solver}
  \label{sdt:algo:quadratic}
  \begin{algorithmic}[1]
  \renewcommand{\algorithmicrequire}{\textbf{Input:}}
  \renewcommand{\algorithmicensure}{\textbf{Output:}}
  \REQUIRE $q[n]$, $h[n]$, $F$, $s$
  \ENSURE  $\phi$
  \\ \textit{Initialization}
   \STATE {$\phi = s \cdot \infty$}
   \STATE {$a = 0$; $b = 0$; $c = -1/F^2$}
  \\ \textit{Process}
   \FOR {$l = 0$ to $n-1$}
   \IF {($s \cdot \phi < s \cdot q[l]$)}
   \STATE break
   \ENDIF
   \STATE {$a \mathrel{+}= 1~/~h[l]^2$}
   \STATE {$b \mathrel{-}= 2 \cdot q[l]~/~h[l]^2 $}
   \STATE {$c \mathrel{+}= q[l]^2~/~h[l]^2 $}
   \STATE {$\phi_1, \phi_2 = \text{SolveQuadratic}(a, b, c)$}
   \STATE {$\phi = s \cdot \max(s \cdot \phi_1, s \cdot \phi_2)$}
   \ENDFOR
  \RETURN $u$
  \end{algorithmic}
\end{algorithm}

\subsubsection{Solving the Equation}
The general procedure of Zhao~\cite{zhao2005fast,zhang2006high} is used to solve Equations~\ref{sdt:eqn:godunov_positive} and~\ref{sdt:eqn:godunov_negative}. 
We summarize the procedure for the positive embedding first.

Consider Equation~\ref{sdt:eqn:godunov_positive} with generalized constants:
\begin{equation}
  \left[\left(\frac{\phi - q_1}{h_1}\right)^+\right]^2 + \left[\left(\frac{\phi - q_2}{h_2}\right)^+\right]^2 + \ldots = \frac{1}{F^2}
\end{equation}
where $q$ is an array of values and $h$ is the array of sample spacing.
There will be a solution $\phi$ where values of $q$ are larger than $\phi$, making the $(\cdot)^+$ operator zero.
The idea is to iteratively solve the quadratic until such a condition is reached or the end of the array is reached.

Practically, this is solved in the following manner.
First, sort the array of $q$'s from smallest to largest, taking care to apply the sorting key to the spacing array as well.
Then, quadratics are iteratively solved until the stopping condition is met.
This is summarized in Algorithm~\ref{sdt:algo:quadratic}.
Note that this algorithm applies to arbitrary dimensions.

The only difference when solving for the negative embedding is that the $q$'s are sorted in decreasing instead of increasing order.
Algorithm~\ref{sdt:algo:quadratic} can then be applied directly.

\subsubsection{Sweeping}
Gauss-Seidel iterations are performed to iteratively improve the embedding function.
The solver sweeps over the image and solves Equation~\ref{sdt:eqn:godunov_positive} if the embedding is positive and Equation~\ref{sdt:eqn:godunov_negative} if the embedding is negative.
However, due to the causality condition, if the sweeping order is alternated, the solution can propagate along the sweeping path.
This is done by sweeping over the image $2^d$ times per iteration, where $d$ is the dimensionality of the image:
\begin{eqnarray*}
  (1): i = 1 \ldots I, j = 1 \ldots J \\
  (2): i = 1 \ldots I, j = J \ldots 1 \\
  (3): i = I \ldots 1, j = 1 \ldots J \\
  (4): i = I \ldots 1, j = J \ldots 1
\end{eqnarray*}
where $I$, $J$ are the number of voxels in the i\textsuperscript{th} and j\textsuperscript{th} directions in a 2D image.
This can be extended to arbitrary dimensions.

\subsubsection{High Order Scheme}
The scheme is now made high order.
Resolution comes from the accuracy of the finite difference approximation. 
We use the approach of Zhao~\cite{zhang2006high} and weighted essentially non-oscillatory (WENO) schemes~\cite{liu1994weighted}.
The idea is to replace $\phi_{min}$ and $\phi_{max}$ in Equations~\ref{sdt:eqn:godunov_positive} and \ref{sdt:eqn:godunov_negative} with a term that makes the finite difference approximation high order.

\begin{figure}
  \centering
  \includegraphics[width=\linewidth]{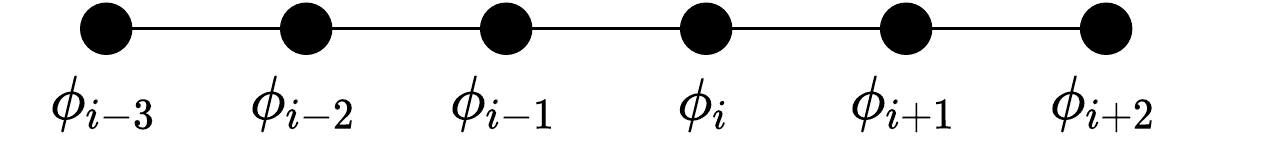}%
  \caption{Stencil for computing the left WENO derivative, $\phi_-$.}
  \label{sdt:fig:weno_stencil}
\end{figure}

Consider the left biased stencil in Figure~\ref{sdt:fig:weno_stencil}.
The derivative is estimated with a fifth order WENO scheme~\cite{jiang2000weighted}.
The weights are described in the original paper.
This process is repeated for the right neighbour, taking the right biased stencil.
Equation~\ref{sdt:eqn:sgnmin} is used to select between the left and right biased estimates.
Let $\phi_-$ be the left derivative and $\phi_+$ be the right derivative in the i\textsuperscript{th} direction.
The upwind derivative is taken as:
\begin{equation}
  q_i = s \cdot \min\left(s (\phi - h_i \phi_-), s (\phi + h_i \phi_+)\right)
\end{equation}

\subsubsection{Boundary Conditions}
While this is a boundary value problem, the edge of the computational domain needs to be handled.
We do a linear extrapolation for the stencil points outside the domain.
An alternative option would be to modify the stencil size based on which neighbors are present~\cite{chopp2001some}, but this leads to considerable computational burden requiring multiple switches at each iteration.

\subsubsection{Preventing Sign Change}
\label{sdt:subsubsec:sign}
We wish to ensure the recovery condition of Equation~\ref{sdt:eqn:recovery} is preserved.
This is achieved by not letting the sign of the embedding change.
Let $\phi^{n+1}$ be the next solution and $\tilde{\phi}^{n+1}$ be the candidate solution from the quadratic solver.
The update equation is:
\begin{equation}
  \phi^{n+1} = s\max(s\tilde{\phi}, \epsilon)
\end{equation}
where $\epsilon$ is a small constant, taken to be the IEEE 754 double precision machine epsilon ($\approx 2.22 \cdot 10^{-16}$).
This small constant prevents the solution from being exactly zero.


\subsubsection{Handling Uniqueness}
Whichever voxel in the narrowband is visited first in the image will likely be assigned $\epsilon$ on the first pass, resulting in the embedding being shifted towards that point.
To help alleviate this, after each $2^d$ sweeps, the embedding is shifted by a constant.
\begin{eqnarray}
  \phi_{lower} &=& \max_{x \in \Omega^-}(\phi(x)) \\
  \phi_{upper} &=& \min_{x \in \Omega^+}(\phi(x)) \\
  \phi &\leftarrow& \phi - \frac{\phi_{lower} + \phi_{upper}}{2}
\end{eqnarray}
We find that this helps with convex shapes such as a sphere, but general shapes will simply have $\phi_{lower} = -\epsilon$ and $\phi_{upper} = \epsilon$.
Since there are no additional constraints on the problem, a unique solution cannot be found, but the solution does satisfy the problem definition (Section~\ref{sdt:sec:definition}).
This is akin to the case in convex optimization, where many problems have no unique solution but many optimal solutions.

\subsubsection{Convergence}
Since a weighted oscillatory scheme is used, oscillations are present around shocks, and a monotone condition cannot be enforced.
Error is measured each update as the absolute value difference between the last and next solution:
\begin{equation}
  \label{sdt:eqn:error}
  E = \frac{1}{2^d N}\sum_i^{2^d} \sum_{x \in \Omega} | \phi^{n+i}(x) - \phi^{n+i+1}(x) |
\end{equation}
where $N$ is the number of voxels in the image and $|\cdot|$ is the absolute value function.
This can be interpreted as the average $\ell^1$ norm at each voxel.
Convergence is assessed by the difference in error between two iterations.
\begin{equation}
  E < \delta
\end{equation}
where $\delta$ is the convergence measure.
A maximum iteration number is also used.
The convergence measure is a departure from the convergence measure of Zhao~\cite{zhao2007parallel} avoiding the need to store the last iteration image, allowing the filter to be run in-place with a reduced memory requirement.

\subsection{Complexity}
The image can be modified in-place, so the space complexity is the number of voxels in the image.
Similarly, the time complexity is $\mathcal{O}(N)$ where $N$ is the number of voxels.
However, the image needs to be swept over $2^d$ times per iteration where $d$ is the dimensionality.
Furthermore, since the monotone condition cannot be enforced, multiple passes are needed over the image.
While the algorithm is linear, it has a large time complexity constant.
The major limitation is that the algorithm cannot be parallelized.
Since monotonicity cannot be enforced, the sweeping directions cannot be combined if they are solved on different threads~\cite{zhao2007parallel}.
The constant can be reduced by only solving the values inside a specified narrowband, avoiding sampling the stencils, computing the WENO derivatives, and solving the quadratic. 
Notably, this problem persists in the FMM where an $\mathcal{O}(N)$ speed up can be achieved in practice if an untidy priority queue is used~\cite{yatziv2006n}.
However, the untidy priority queue cannot be used for high order schemes because quantization of priority causes the embedding to be no more than $\mathcal{O}(h)$ accurate after marching~\cite{yatziv2006n}.

\section{Experiments}
\label{sdt:sec:experiments}
A series of experiments are performed to verify the properties and test the performance of the algorithm.
Since there is no unique solution, experiments measuring the order of accuracy are not obvious to perform.
Similarly, local geometric properties such as mean and Gaussian curvature cannot be assessed for order of accuracy since the subvoxel shift permits a change in these parameters.
For instance, a sphere of radius $r$ initialized exactly at the center of a voxel can take on any of the following for mean ($H$) and Gaussian ($K$) curvature:
\begin{eqnarray}
  H = \left\{ \frac{1}{r + a} \middle| -h/2 < a < h/2\right\} \\
  K = \left\{ \frac{1}{(r + a)^2} \middle| -h/2 < a < h/2\right\}
\end{eqnarray}


\begin{figure*}
  \centering
  \begin{tabular}{cccccc}
    \subfloat[Initialization]{
      \includegraphics[width=0.13\linewidth]{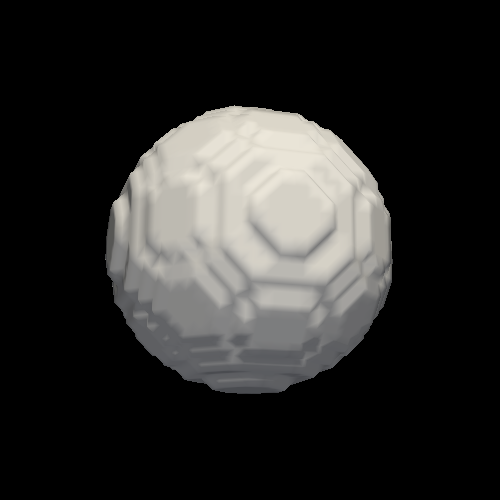}%
      \label{sdt:fig:convergence:I:0}
    } & 
    \subfloat[1 Iteration]{
      \includegraphics[width=0.13\linewidth]{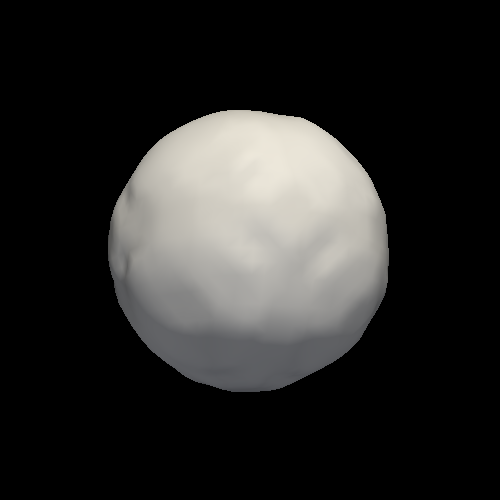}%
      \label{sdt:fig:convergence:I:1}
    } & 
    \subfloat[2 Iteration]{
      \includegraphics[width=0.13\linewidth]{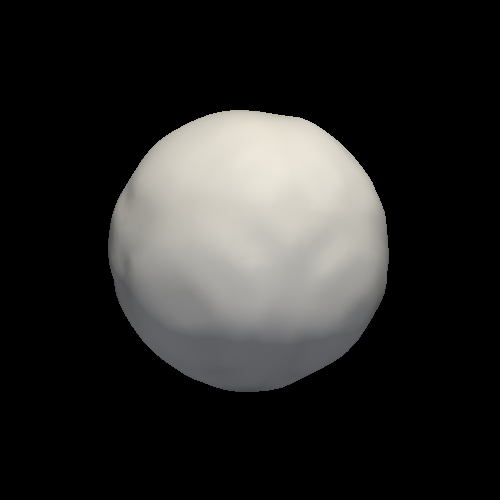}%
      \label{sdt:fig:convergence:I:2}
    } &
    \subfloat[10 Iteration]{
      \includegraphics[width=0.13\linewidth]{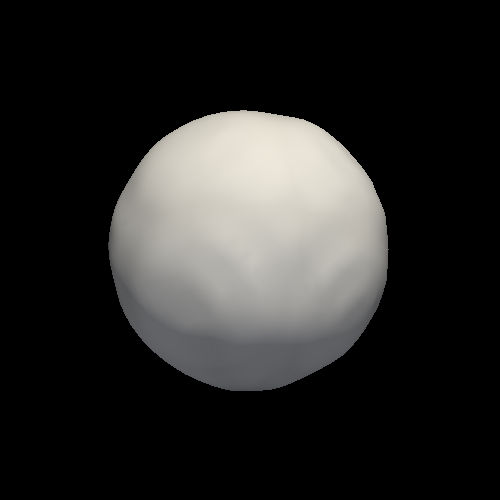}%
      \label{sdt:fig:convergence:I:10}
    } & 
    \subfloat[25 Iteration]{
      \includegraphics[width=0.13\linewidth]{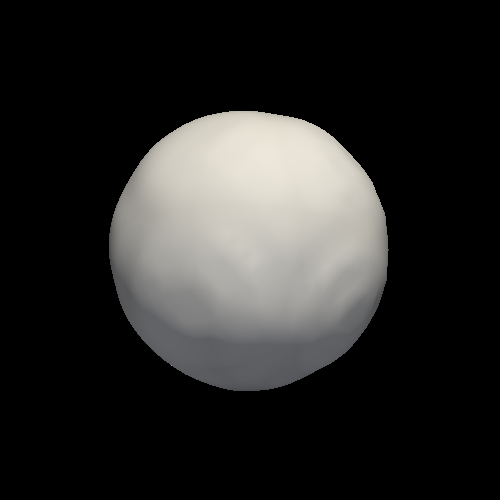}%
      \label{sdt:fig:convergence:I:25}
    } &
    \subfloat[50 Iteration]{
      \includegraphics[width=0.13\linewidth]{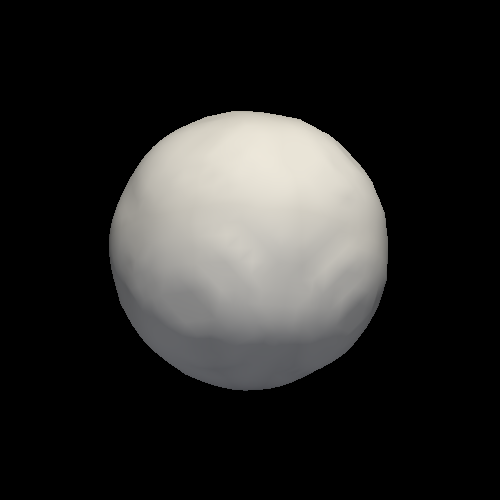}%
      \label{sdt:fig:convergence:I:50}
    }
  \end{tabular}
  \caption{Visualizing the zero isocontour for various iterations ($h = \SI{2}{\milli\meter}$, $r = \SI{25}{mm}$). The first iteration has a notable change while further iterations are subtle.}
  \label{sdt:fig:convergence}
\end{figure*}

\begin{figure}[b]
  \centering
  \includegraphics[width=0.7\linewidth]{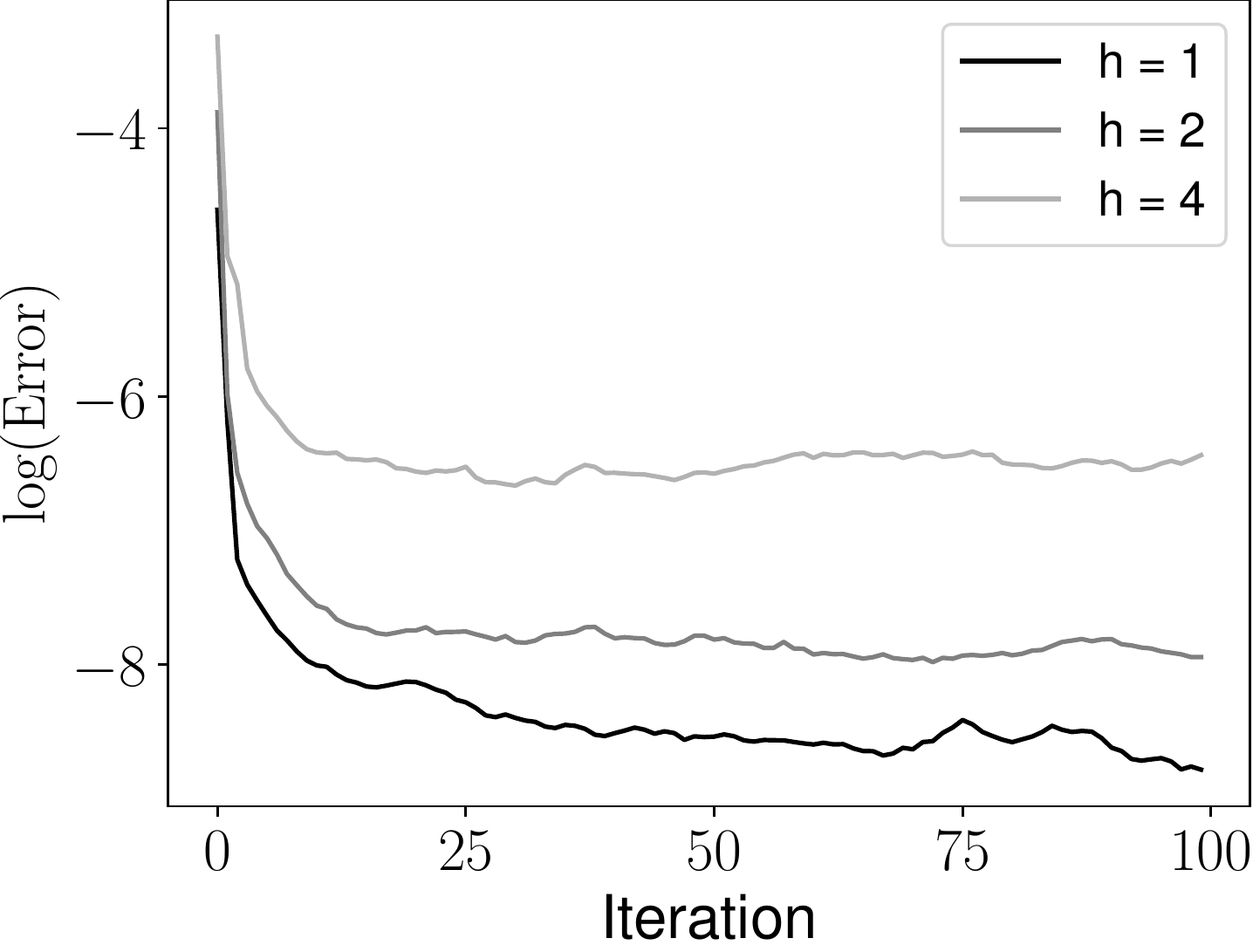}%
  \caption{Log-Error as a function of iterations for a sphere of radius $\SI{25}{\milli\meter}$. Each iteration is $2^d$ sweeps.}
  \label{sdt:fig:error}
\end{figure}

\subsection{Convergence Behavior}
The convergence behavior is assessed empirically.
A three dimensional sphere of radius $\SI{25}{\milli\meter}$ is instantiated on a grid of physical size $\SI{100}{mm}$ with spacings of $h = \left\{4, 2, 1 \right\}\SI{}{\milli\meter}$ resulting in a grid size of $25$, $50$, and $100$.
The ideal signed distance transform is binarized by a threshold at zero and the signed distance transform computed with the proposed method.
100 iterations are performed and the error (Equation~\ref{sdt:eqn:error}) reported at each iteration.
The solution is solved in the $\SI{15}{\milli\meter}$ narrowband.
This avoids including the influence of domain boundary voxels from being measured in the error as they oscillate due to the linear extrapolation.

The zero isocontour is rendered across many iterations in Figure~\ref{sdt:fig:convergence}.
An immediate visual improvement can be seen in a single iteration with very small continued improvement as the iterations increase.
The log-log plot of error as a function of iteration is shown in Figure~\ref{sdt:fig:error}.
The error is not monotone and stabilizes after a few iterations.
The first few iterations very quickly decrease the total error.
Error decreases with decreasing spacing.

\subsection{Order of Accuracy}
\label{sdt:subsec:order_of_accuracy}
Even though no unique solution is available, an attempt is made to measure the order of accuracy.
The order of accuracy is estimated by:
\begin{equation}
  m = \log_2 \left( \frac{\lVert \phi_h - \phi \rVert_p}{\lVert \phi_{h/2} - \phi  \rVert_p} \right)
\end{equation}
where $\phi$ is the ground truth, $\phi_h$ is the solution at spacing $h$, $\phi_{h/2}$ is the solution at spacing $h/2$, and $\lVert \cdot \rVert_p$ is an $\ell^p$ norm.
A three dimensional sphere of radius $\SI{25}{\milli\meter}$ is instantiated on a grid of physical size $\SI{100}{mm}$ with spacings of $h = \left\{8, 4, 2, 1 \right\}\SI{}{\milli\meter}$.
The sphere is binarized, the proposed signed distance transform computed,
and the $\ell^1$ and $\ell^\infty$ norms measured.
Note that $\ell^1$ is the average absolute value between each voxel and not the image norm, while $\ell^\infty$ is the maximum per-voxel error across the image.
Since the embedding can be shifted by a constant, the experiment is performed twice.
First as-is and a second time by finding a constant that minimizes the $\ell^1$ norm and subtracting it from the computed embedding.
The error is only measured in a $\SI{15}{\milli\meter}$ narrowband around the zero level set to avoid errors introduced by linear extrapolation at the boundary.



\begin{table}[b]
  \centering
  \begin{tabular}{c|cccc|cccc}
    \hline
    & \multicolumn{4}{c}{Without Correction} & \multicolumn{4}{c}{With Correction} \\
    \hline
    h & $\ell_1$ & Order & $\ell_\infty$ & Order & $\ell_1$ & Order & $\ell_\infty$ & Order \\
    \hline
    8 & 1.73 & -- & 4.47 & -- & 1.35 & -- & 4.75 & -- \\
    4 & 0.26 & 2.75 & 1.02 & 2.13 & 0.26 & 2.38 & 0.89 & 2.42 \\
    2 & 0.21 & 0.31 & 0.62 & 0.72 & 0.14 & 0.89 & 0.43 & 1.03 \\
    1 & 0.12 & 0.76 & 0.42 & 0.56 & 0.07 & 1.09 & 0.37 & 0.23 \\
   \hline
  \end{tabular}
  \caption{Measuring the order of accuracy in the signed distance map.}
  \label{sdt:table:order}
\end{table}

\begin{figure}[b]
  \centering
  \begin{tabular}{cccc}
    \subfloat[$h = \SI{8}{\milli\meter}$]{
      \includegraphics[width=0.18\linewidth]{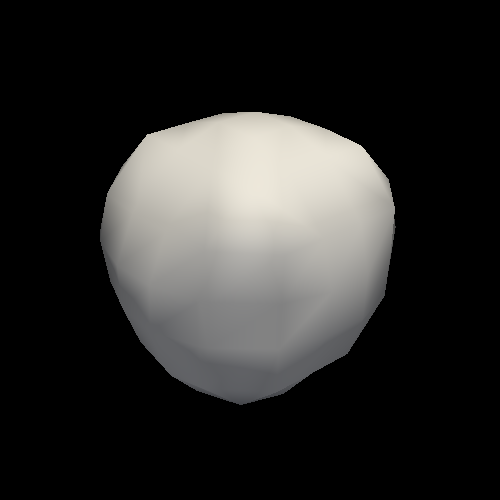}%
      \label{sdt:fig:increasing_r:8}
    } & 
    \subfloat[$h = \SI{4}{\milli\meter}$]{
      \includegraphics[width=0.18\linewidth]{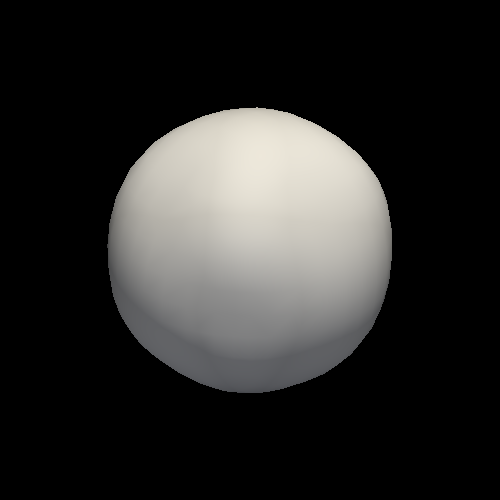}%
      \label{sdt:fig:increasing_r:4}
    } &
    \subfloat[$h = \SI{2}{\milli\meter}$]{
      \includegraphics[width=0.18\linewidth]{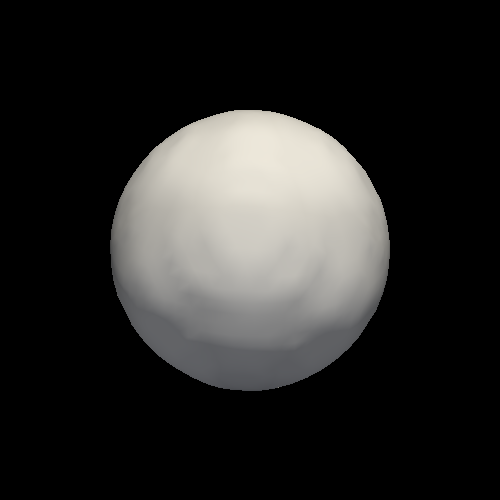}%
      \label{sdt:fig:increasing_r:2}
    } & 
    \subfloat[$h = \SI{1}{\milli\meter}$]{
      \includegraphics[width=0.18\linewidth]{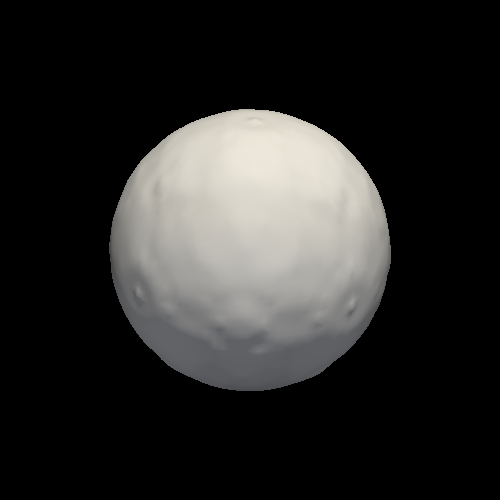}%
      \label{sdt:fig:increasing_r:1}
    }
  \end{tabular}
  \caption{Zero isocontour of a sphere ($r = \SI{25}{\milli\meter}$) binarized and transformed with the proposed method at different image spacing. As the ratio of radius to spacing increases, the rendering approaches a more spherical shape while noise is more obviously.}
  \label{sdt:fig:increasing_r}
\end{figure}

Order of accuracy and error are reported in Table~\ref{sdt:table:order}.
The $\ell^1$ and $\ell^\infty$ are only first-order as is expected from the non-uniqueness of the problem.
In general, the embedding can be smoother (in the $\ell^1$ norm) than suggested by Table~\ref{sdt:table:order}, but will require an extra constraint permitting a unique solution.
However, due to the presence of shocks in the solution, the $\ell^\infty$ norm can be no better than first-order in general.
The degenerate order of accuracy is due to the problem having no unique solution and permitting a sub-voxel shift by a constant on the order of the spacing.
The resulting zero isocontour are visualized in Figure~\ref{sdt:fig:increasing_r}.
At a small ratio of radius to spacing, the surface appears smooth but is not spherical.
As the ratio increases, the surface becomes less distorted but has a presence of noise on the surface.
While being not perfect, these results are still an improvement over the ``exact'' signed distance transform.

\begin{figure*}
  \centering
  \begin{tabular}{ccccccc}
    \subfloat[Truth]{
      \includegraphics[width=0.105\linewidth]{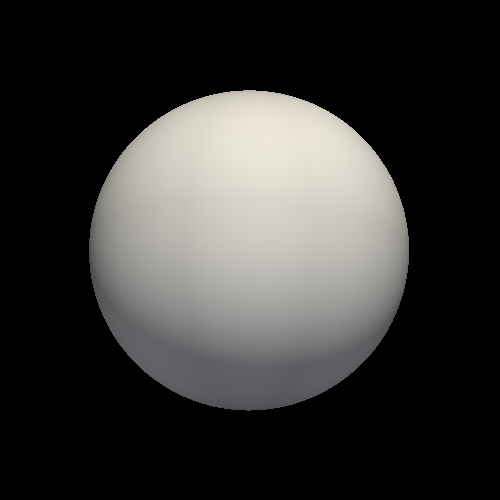}%
      \label{sdt:fig:noise:no:3d}
    } & 
    \subfloat[$m = 0$]{
      \includegraphics[width=0.105\linewidth]{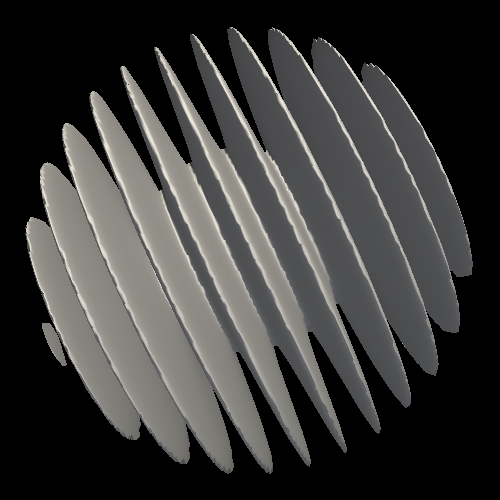}%
      \label{sdt:fig:noise:0:3d}
    } & 
    \subfloat[$m = 1$]{
      \includegraphics[width=0.105\linewidth]{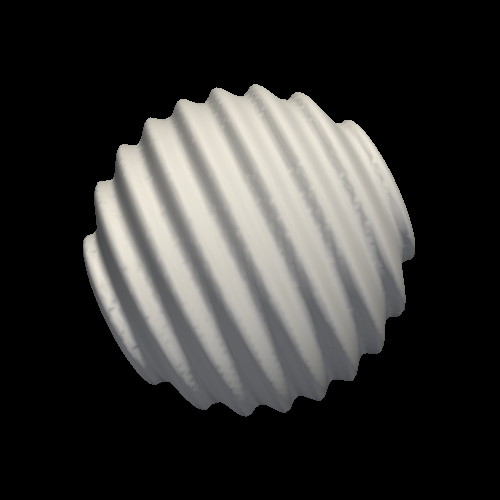}%
      \label{sdt:fig:noise:1:3d}
    } & 
    \subfloat[$m = 2$]{
      \includegraphics[width=0.105\linewidth]{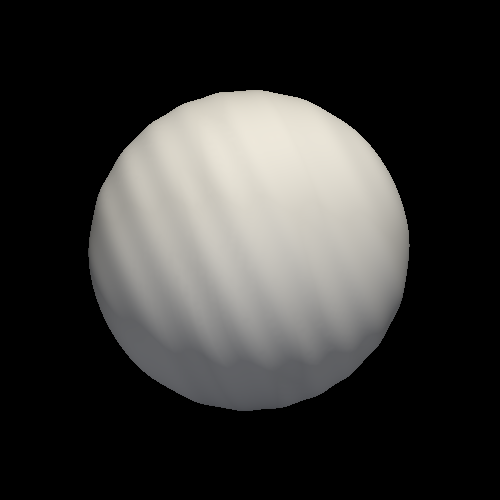}%
      \label{sdt:fig:noise:2:3d}
    } & 
    \subfloat[$m = 3$]{
      \includegraphics[width=0.105\linewidth]{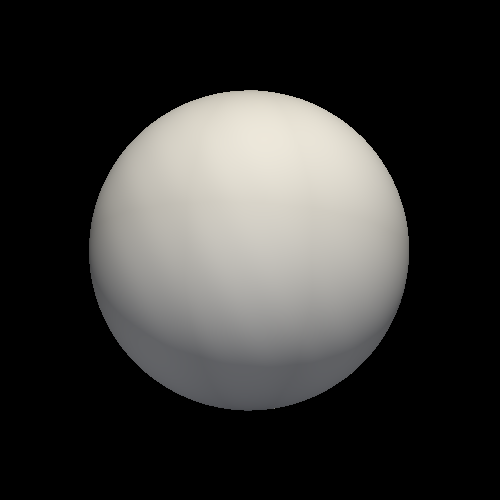}%
      \label{sdt:fig:noise:3:3d}
    } & 
    \subfloat[Exact]{
      \includegraphics[width=0.105\linewidth]{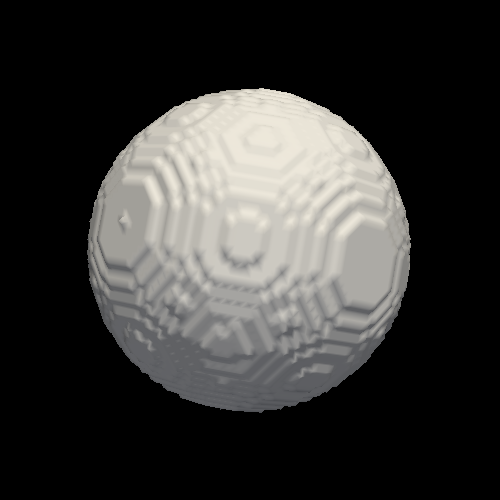}%
      \label{sdt:fig:noise:sdt:3d}
    } &
    \subfloat[\scriptsize Proposed]{
      \includegraphics[width=0.105\linewidth]{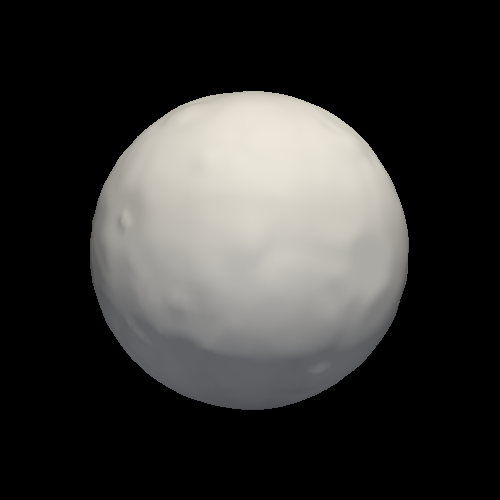}%
      \label{sdt:fig:noise:proposed:3d}
    } \\
    \subfloat[Truth]{
      \includegraphics[width=0.105\linewidth]{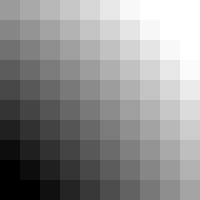}%
      \label{sdt:fig:noise:no}
    } & 
    \subfloat[$m = 0$]{
      \includegraphics[width=0.105\linewidth]{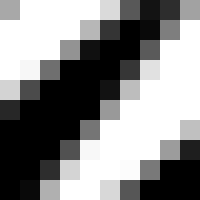}%
      \label{sdt:fig:noise:0}
    } & 
    \subfloat[$m = 1$]{
      \includegraphics[width=0.105\linewidth]{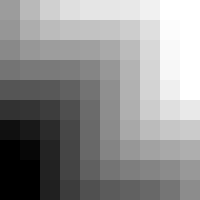}%
      \label{sdt:fig:noise:1}
    } & 
    \subfloat[$m = 2$]{
      \includegraphics[width=0.105\linewidth]{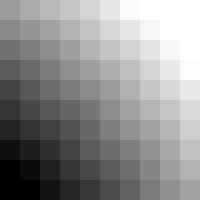}%
      \label{sdt:fig:noise:2}
    } & 
    \subfloat[$m = 3$]{
      \includegraphics[width=0.105\linewidth]{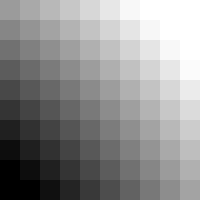}%
      \label{sdt:fig:noise:3}
    } & 
    \subfloat[Exact]{
      \includegraphics[width=0.105\linewidth]{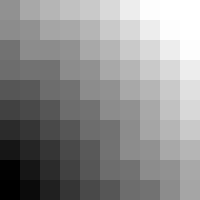}%
      \label{sdt:fig:noise:sdt}
    } &
    \subfloat[\scriptsize Proposed]{
      \includegraphics[width=0.105\linewidth]{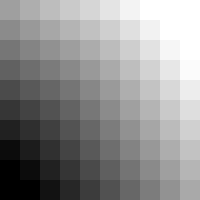}%
      \label{sdt:fig:noise:proposed}
    }
  \end{tabular}
  \caption{Effect of additional noise on a sphere ($h = 0.1$, $r = 2.5$). While the case of $m=0$ appears scaled and to face a different direction relative to the other cases, this effect is caused by the noise amplitude being 4 times larger than the radius. Zero isocontours are rendered above and a 2D slice is extracted on bottom at the upper corner of the sphere. 2D slices are visualized in the dynamic range $\SI{-0.5}{mm}$ to $\SI{0.5}{mm}$.}
  \label{sdt:fig:noise}
\end{figure*}

\subsection{Comparison to Adding Noise}
Noise of a prescribed order of accuracy is added to an analytic signed distance map to compare against the proposed method~\cite{coquerelle2016fourth}.
The noise has a frequency of $\frac{1}{10h}$ and amplitude that depends on the spacing to the power of the order.
\begin{equation}
  \tilde{\phi} = \phi + h^m \sin\left(\frac{2\pi}{10h}\left(x+y+z\right)\right)
\end{equation}
As in Section~\ref{sdt:subsec:order_of_accuracy}, $m$ is the order of accuracy and $h$ is the sample period.
This is an empirical method of adding noise of prescribed order of accuracy to the embedding.
As the frequency depends on the sample spacing:
\begin{equation}
  \frac{\partial \tilde{\phi}}{\partial x} = \frac{\partial \phi}{\partial x} + \frac{2\pi}{10} h^{m-1} \cos\left(\frac{2\pi}{10h}\left(x+y+z\right)\right)
\end{equation}
The frequency is a tenth of sample spacing to avoid aliasing while still having a diminishing amplitude with derivative.

The ideal signed distance transform of a sphere with radius $r=\SI{2.5}{mm}$ is instantiated on a $\SI{10}{mm}$ edge length grid with spacing $h=\SI{0.1}{mm}$.
The sphere is instantiated with added noise, the exact signed distance transform, and the proposed method are shown in Figure~\ref{sdt:fig:noise}.
The exact signed distance transform exhibits a visual smoothness between first and second order accurate.
The proposed technique exhibits a visual smoothness between second and third order accurate.
The proposed method is not as accurate as the WENO stencil (fifth order).
A degenerate accuracy is believed to have two causes.
First, the non-uniqueness of the problem permits a sub-voxel shift in the embedding.
Instead of being an additive constant, the proposed method produces local shifts in the rendered sphere.
This effect is seen more clearly in Figures~\ref{sdt:fig:result:proposed} and \ref{sdt:fig:increasing_r:8}.
Second, the initialization is only first-order accurate in the narrowband due to the quantization artifact of the exact signed distance transform.
The original fast sweeping method~\cite{zhao2005fast} relies on a high order initialization.
It was expected that over many iterations the narrowband would further improve, but Figure~\ref{sdt:fig:convergence} is showing a slow rate of convergence.

\subsection{Application to Real Data}
The method is applied to real data to assess the applicability.
One fourth lumbar vertebra and proximal femur were manually segmented from a clinical computed tomography dataset with slice thickness $\SI{0.625}{\milli\meter}$ and in-plane spacing $\SI{0.699}{\milli\meter}$.
The segmentations contain slice wise contouring artifacts which were cleaned using morphological opening and closing operators:
\begin{equation}
  J = \text{Open}\left(\text{Close}\left(I, S\right), S\right)
\end{equation}
where $I$ is the manually segmented data with slice wise contouring artifacts, $S$ is a structuring element, and $\text{Open}(\cdot)$ and $\text{Close}(\cdot)$ are morphological opening and closing.
A nice property of this surface cleaning technique is that the resulting image has the local smoothness of the structuring element.
A spherical structuring element is used here with a prescribed radius $r$ such that the maximum mean curvature in the image is $1/r$.
One could choose instead to open then close resulting in an image with more holes along its surface.
However, as we are working with generally convex biological images, we want the surface to not have these holes.

The images were cleaned with a spherical structuring element of radius one voxel and embedded using the exact signed distance transform and high-order signed distance transform.
The zero isocontour is extracted using marching cubes~\cite{lorensen1987marching}, vertices interpolated into the signed distance field, and the mean and Gaussian curvature measured~\cite{besler2021morph}.
Having the mean and Gaussian curvature at each vertex allows a histogram to be generated for an image.
Additionally, the Euler-Poincar\'e characteristic, $\chi$, was measured using the Gauss-Bonnet theorem~\cite{besler2021morph}.

\begin{figure*}
  \centering
  \begin{tabular}{ccc}
    \subfloat[Exact L4]{
      \includegraphics[width=0.26\linewidth]{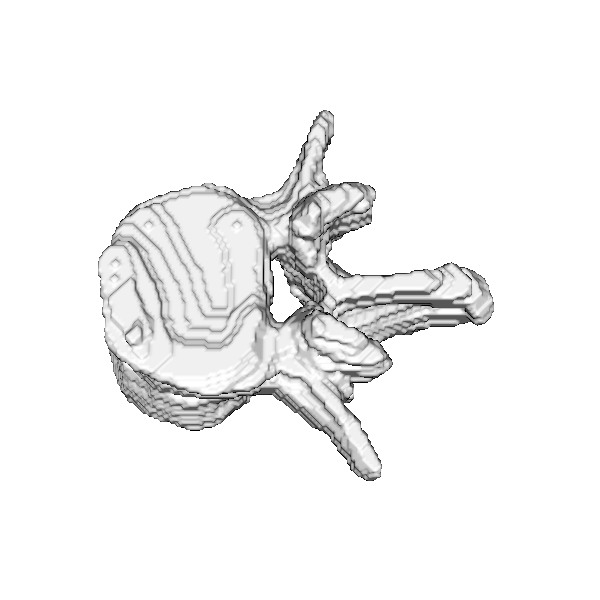}%
      \label{sdt:fig:surface:l4:exact:surface}
    } & 
    \subfloat[Mean Curvature]{
      \includegraphics[width=0.26\linewidth]{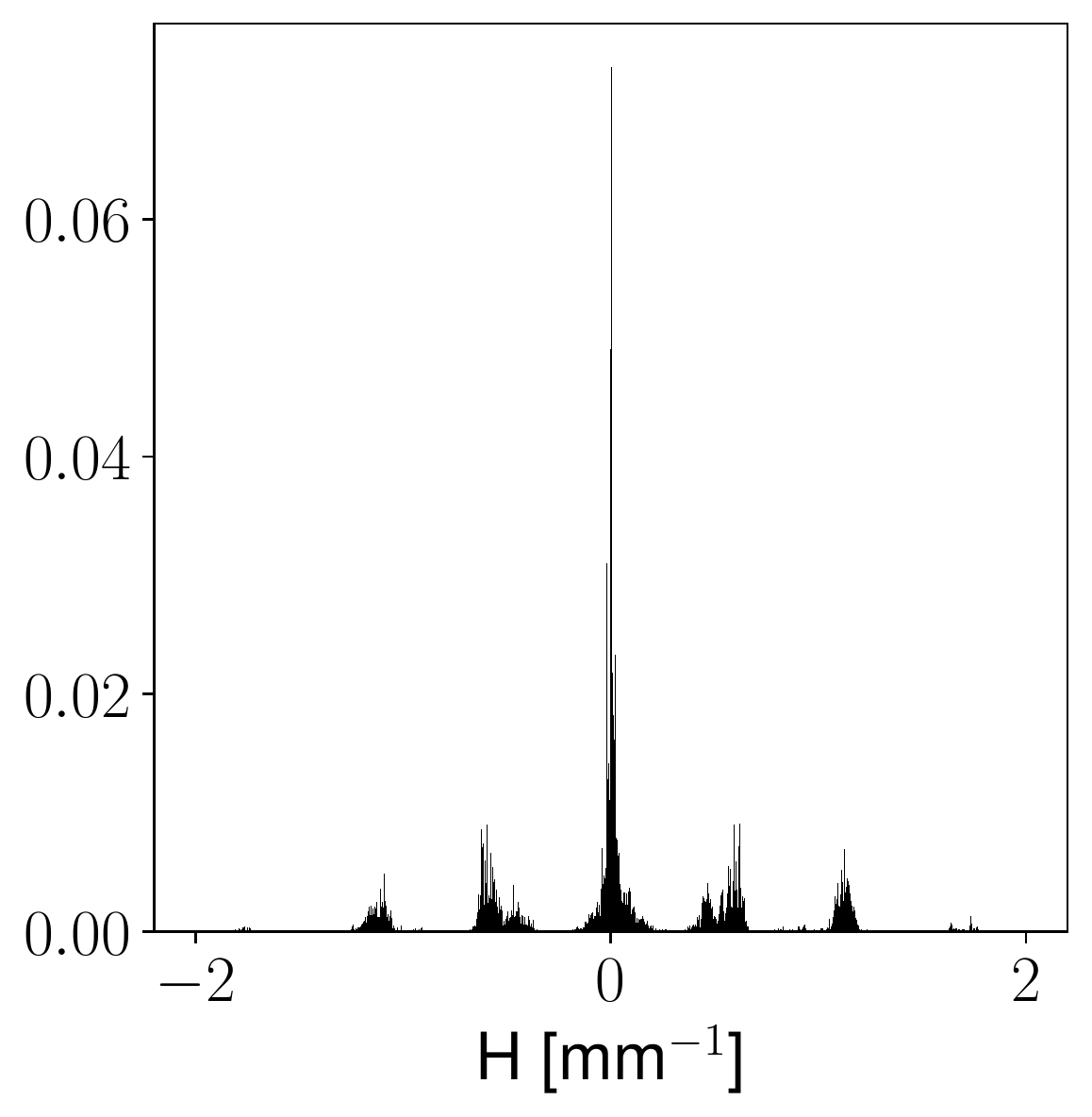}%
      \label{sdt:fig:surface:l4:exact:mean}
    } & 
    \subfloat[Gaussian Curvature]{
      \includegraphics[width=0.26\linewidth]{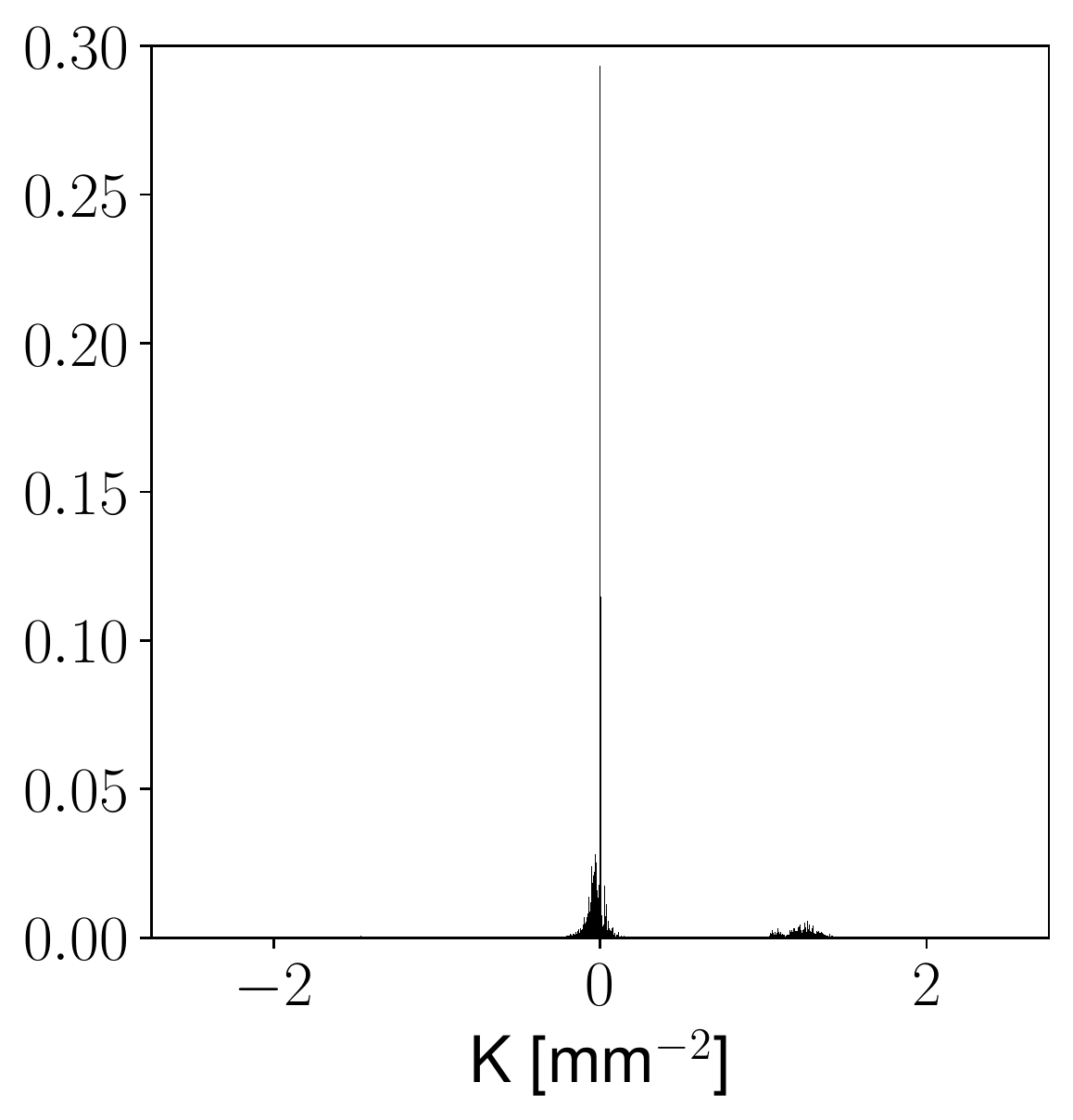}%
      \label{sdt:fig:surface:l4:exact:gauss}
    } \\
    \subfloat[High-Order L4]{
      \includegraphics[width=0.26\linewidth]{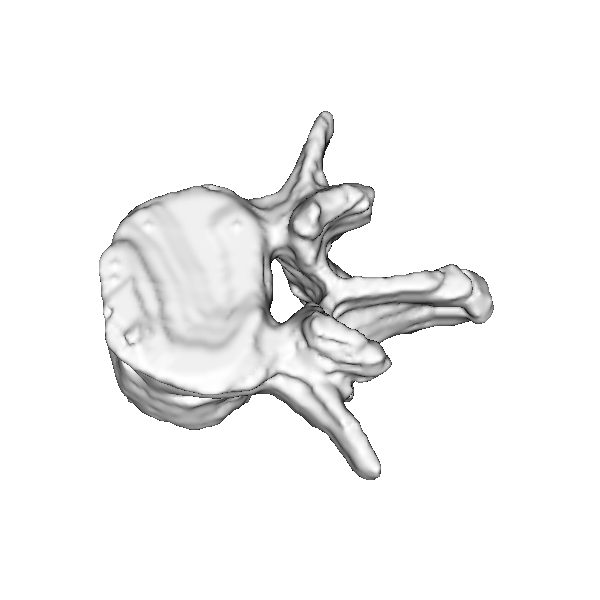}%
      \label{sdt:fig:surface:l4:proposed:surface}
    } & 
    \subfloat[Mean Curvature]{
      \includegraphics[width=0.26\linewidth]{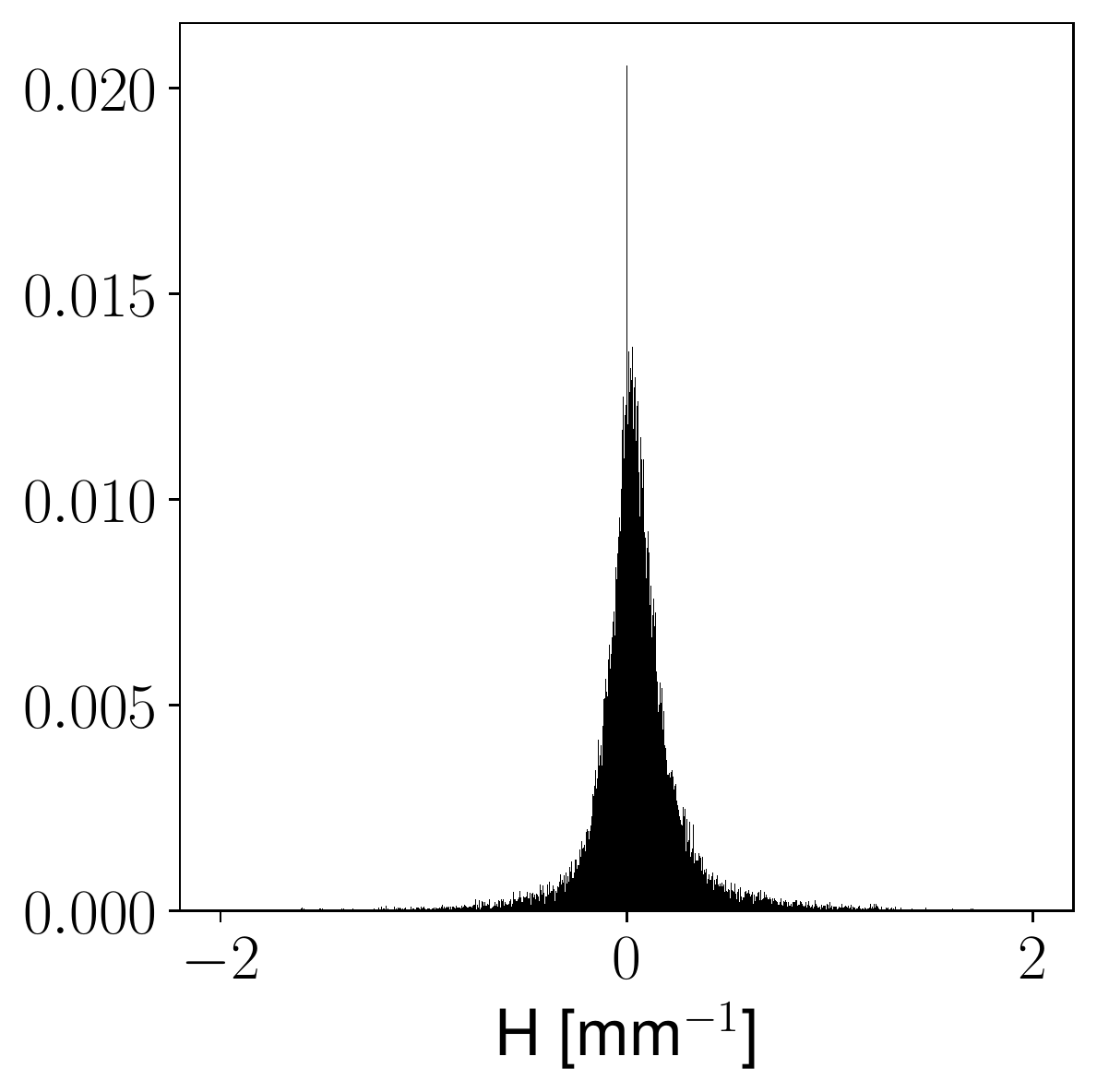}%
      \label{sdt:fig:surface:l4:proposed:mean}
    } & 
    \subfloat[Gaussian Curvature]{
      \includegraphics[width=0.26\linewidth]{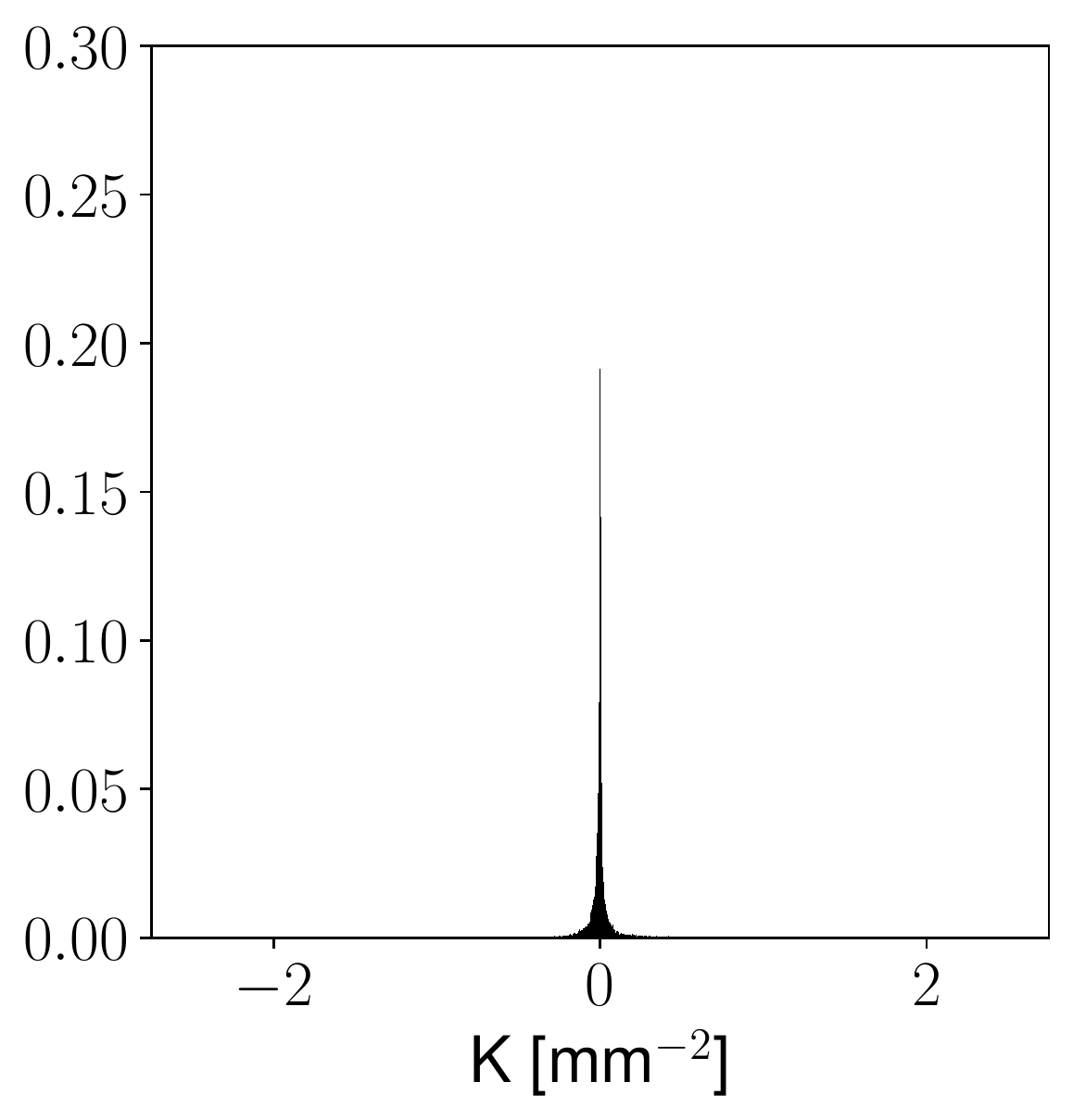}%
      \label{sdt:fig:surface:l4:proposed:gauss}
    } \\
    \subfloat[Exact Femur]{
      \includegraphics[width=0.26\linewidth]{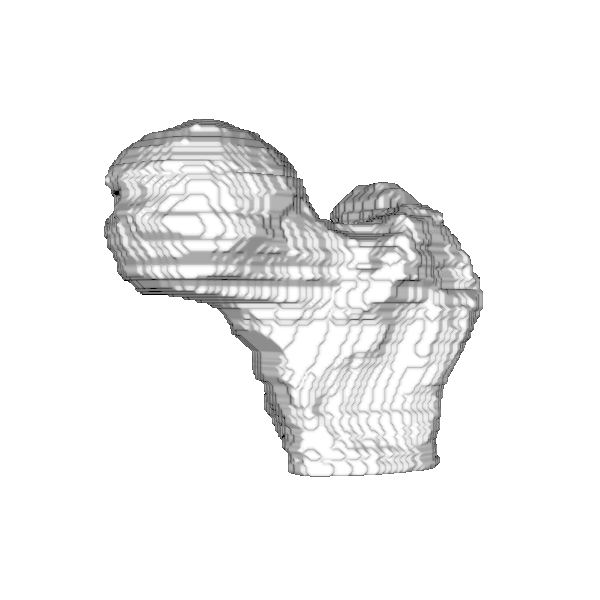}%
      \label{sdt:fig:surface:femur:exact:surface}
    } & 
    \subfloat[Mean Curvature]{
      \includegraphics[width=0.26\linewidth]{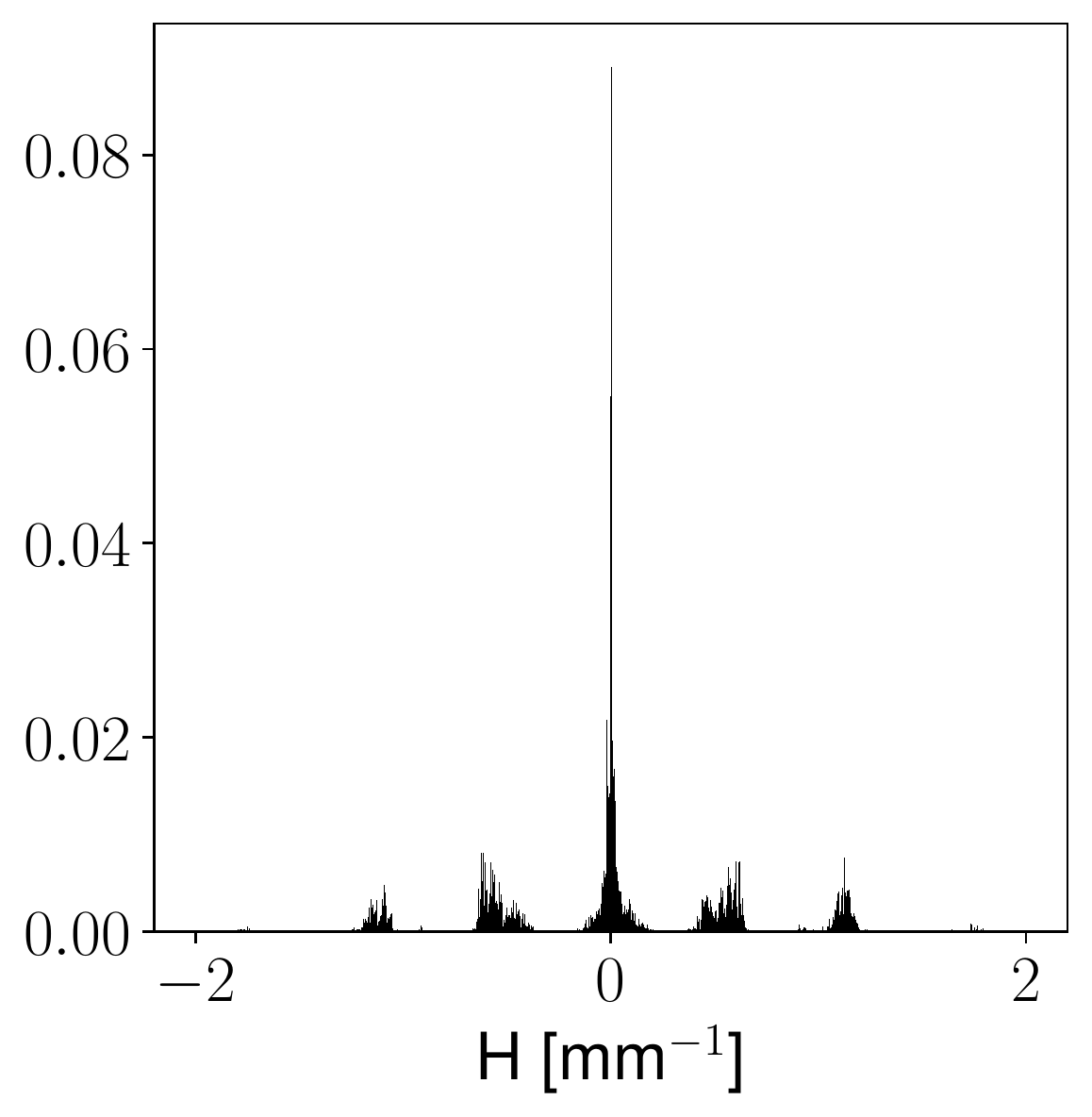}%
      \label{sdt:fig:surface:femur:exact:mean}
    } & 
    \subfloat[Gaussian Curvature]{
      \includegraphics[width=0.26\linewidth]{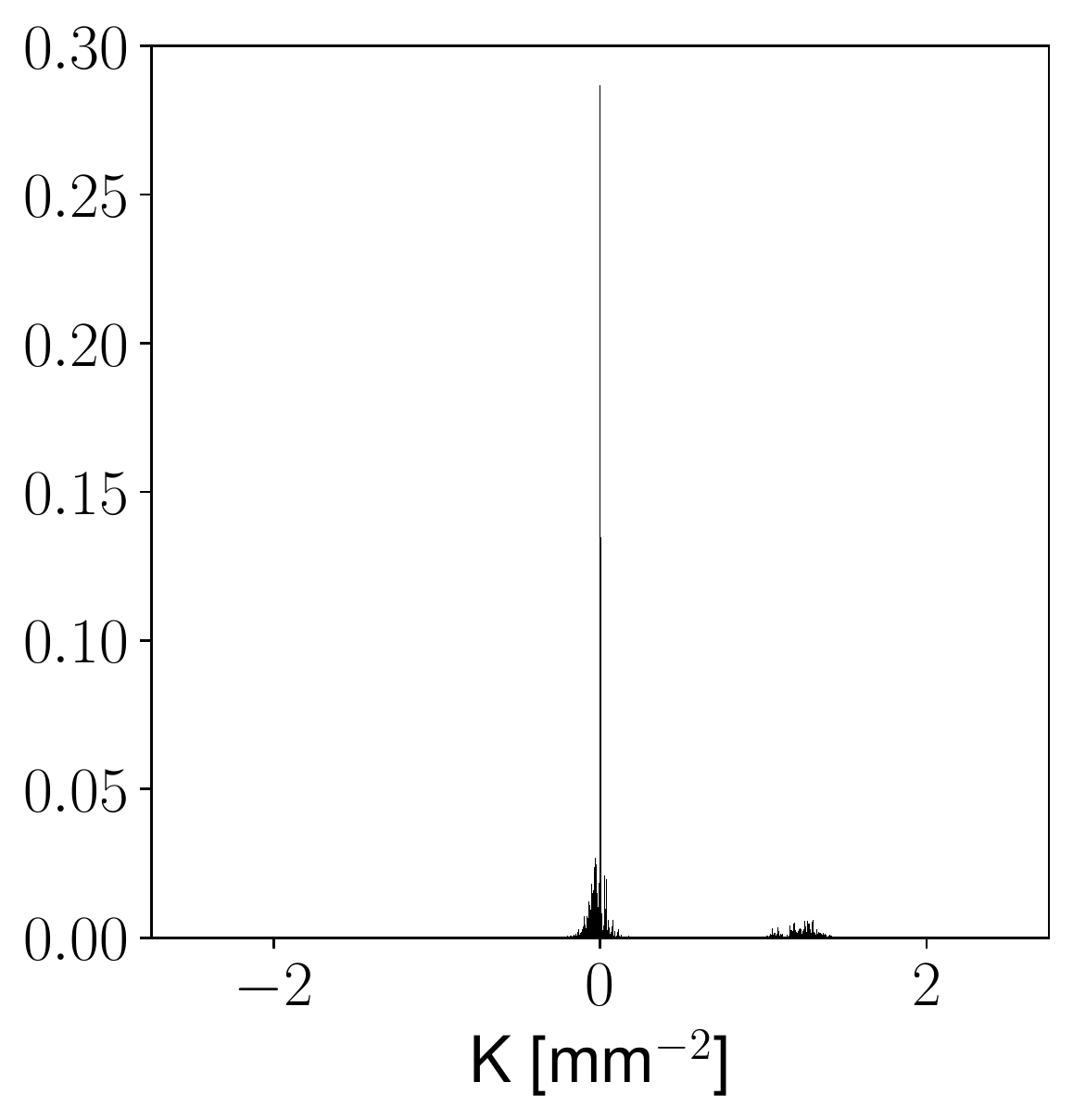}%
      \label{sdt:fig:surface:femur:exact:gauss}
    } \\
    \subfloat[High-Order Femur]{
      \includegraphics[width=0.26\linewidth]{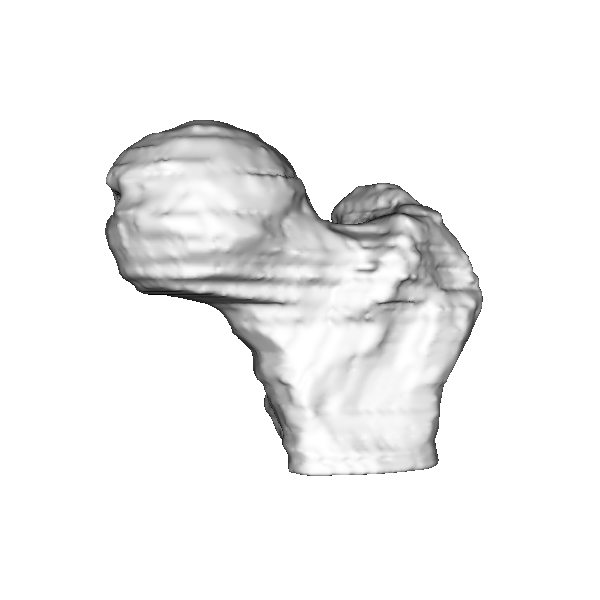}%
      \label{sdt:fig:surface:femur:proposed:surface}
    } & 
    \subfloat[Mean Curvature]{
      \includegraphics[width=0.26\linewidth]{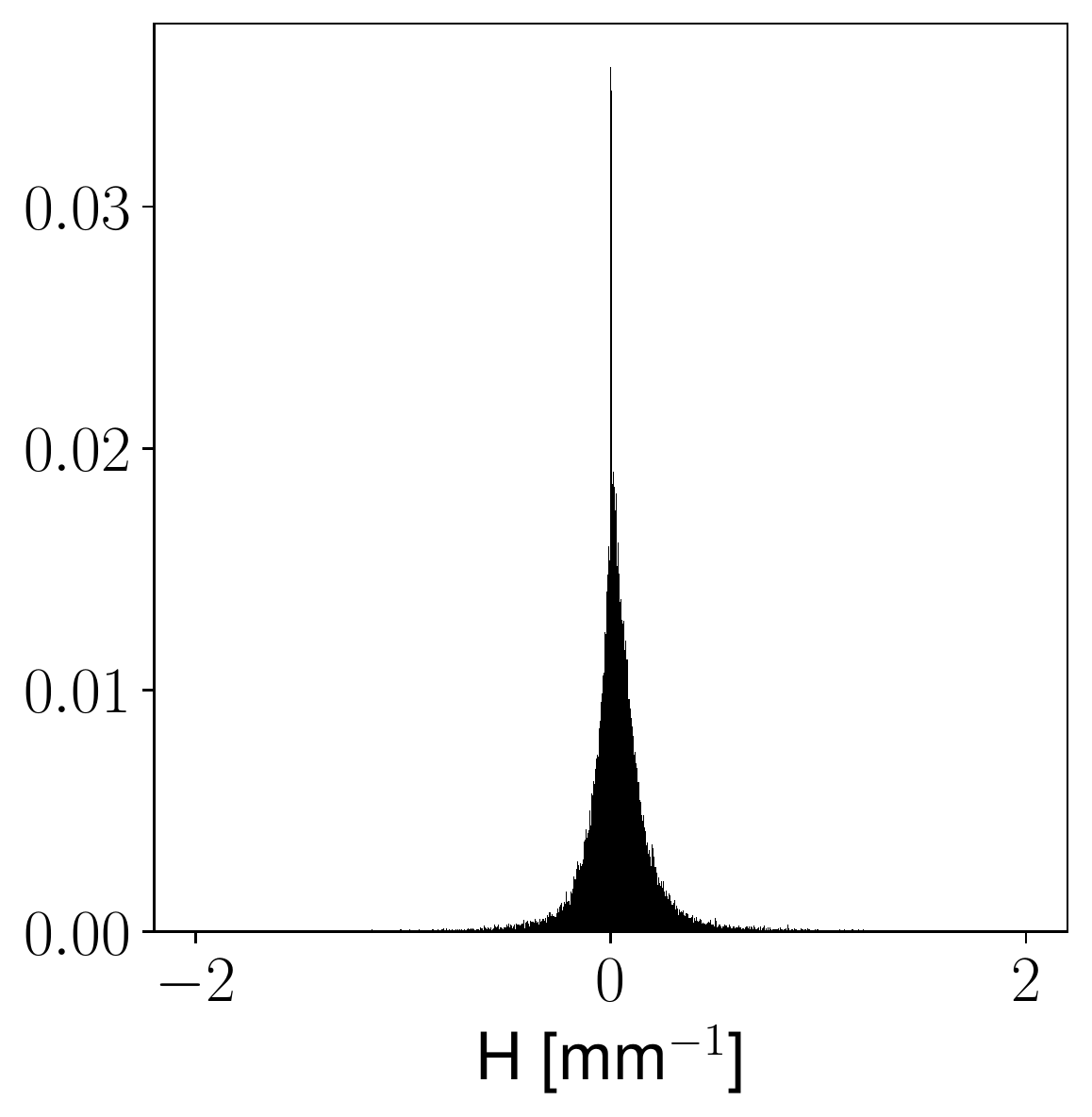}%
      \label{sdt:fig:surface:femur:proposed:mean}
    } & 
    \subfloat[Gaussian Curvature]{
      \includegraphics[width=0.26\linewidth]{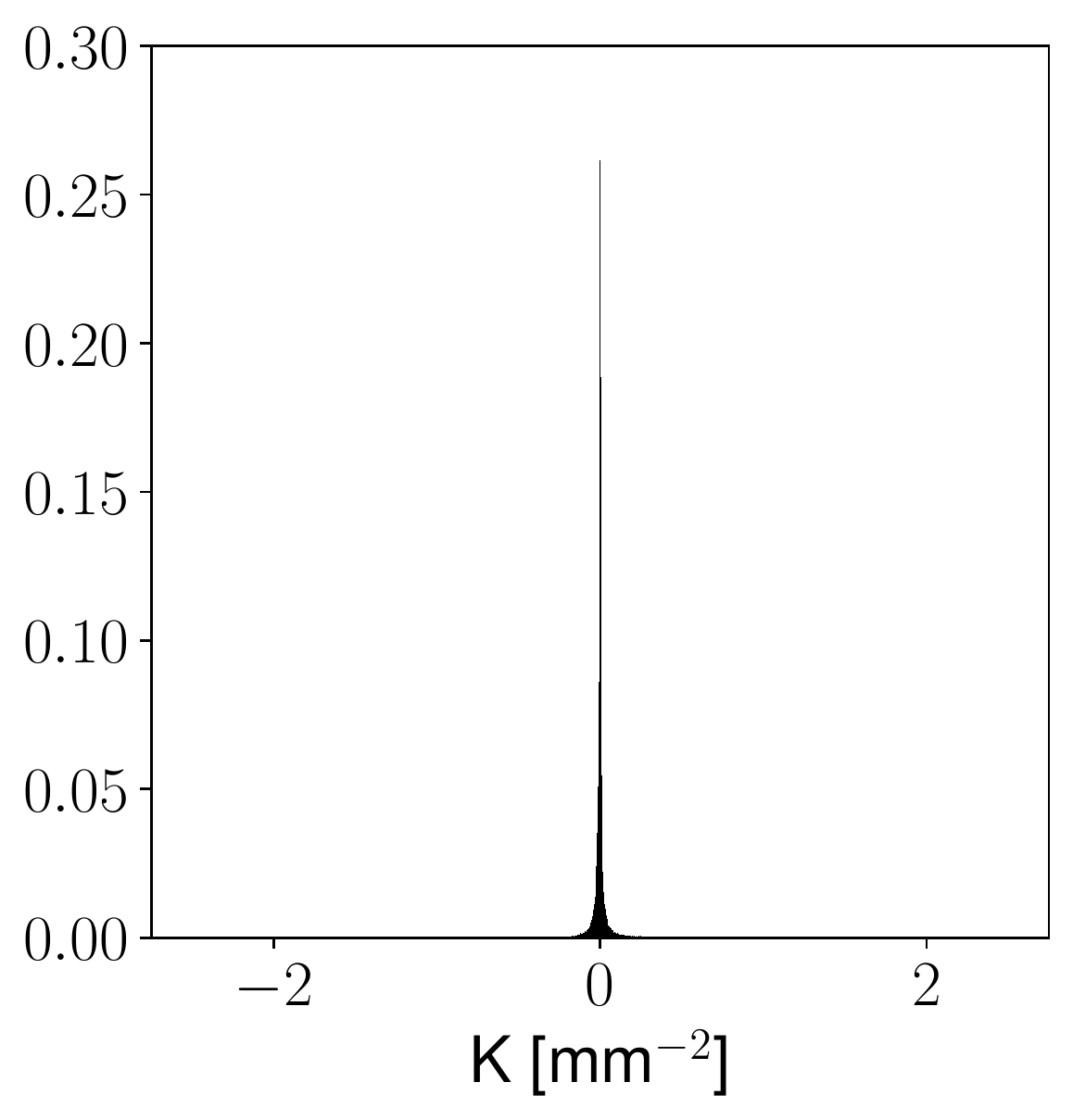}%
      \label{sdt:fig:surface:femur:proposed:gauss}
    }
  \end{tabular}
  \caption{Rendering and curvature histograms for the exact signed distance transform and high-order signed distance transform for a fourth lumbar vertebra and proximal femur. Quantization artifacts are seen in the mean and Gaussian curvature histograms. The high-order signed distance transform is visibly much smoother given the exact version, even though the recovery condition is met in both.}
  \label{sdt:fig:surface}
\end{figure*}

The histograms and meshes are visualized for the vertebra and femur in Figure~\ref{sdt:fig:surface}.
The exact signed distance transform is less smooth and has artifacts in the mean and Gaussian curvature histograms, well described elsewhere~\cite{besler2020artifacts}.
Of note, the recovery condition is met in both cases (the difference in Heaviside gives an image of all zeros).
Slice wise contouring artifacts are still seen in the high-order distance transform, but effects of sampling are removed.
The measured Euler-Poincar\'e characteristic was 18.57 (exact) and 1.18 (high-order) for the vertebrae and 20.24 (exact) and 3.6 (high-order) for the femur.
The ground truth values are 0 for the vertebrae and 2 for the femur.
While the proposed high-order signed distance transform is greatly improved compared to the exact signed distance transform, it is still not accurate enough to be used for measuring the Euler-Poincar\'e characteristic.

\section{Discussion}
\label{sdt:sec:discussion}
The problem of computing a high order signed distance transform of a binary image is presented as a transform satisfying the following criterion: recovery the original transform (Equation~\ref{sdt:eqn:recovery}), satisfy the Eikonal equation (Equation~\ref{sdt:eqn:eikonal}), and be of order greater than unity (Equation~\ref{sdt:eqn:smoothness}).
Based on the results of this paper and experience developing these algorithms, the authors believe there does not exist a transform $T$ that satisfies the recovery equation, the Eikonal equation, and is of order higher than unity in a finite sized narrowband for arbitrary binary images $I$. 
We do not have the analytic abilities to prove this postulate, only to give the reasoning of Section~\ref{sdt:subsec:suface_shocks}.
A prototype image for testing this postulate would be a checkerboard, where every voxel is opposite of the last.
If a checkerboard image could be transformed and satisfy the three criteria, this would be reasonable evidence that the postulate is incorrect.
However, if the postulate is true, the question becomes what new criterion is needed.
We expect that the criterion will manifest as surface editing (smoothing) that relaxes the recovery condition.

The issue of computing a high order signed distance transform originates in the fact that the binary image is sampled (spatially discretized)~\cite{besler2020artifacts}.
An analogous but different issue arose in scale space theory where the optimal kernel for continuous functions was the Gaussian~\cite{babaud1986uniqueness} while the optimal kernel for sampled functions was an exponential multiplied by a modified Bessel function~\cite{lindeberg1990scale}.
While existence and uniqueness proofs exist for the continuous Eikonal equation based on the viscous solution~\cite{crandall1983viscosity}, the signed distance transform of sampled signals are a different problem not affording the same analysis.
Specifically, this problem is the initialization of the narrowband.

The proposed method was termed the ``signed fast sweeping method'' based on adding a sign condition to Zhao's fast sweeping method~\cite{zhao2005fast}.
References to a signed fast sweeping method can be found in literature~\cite{oberhuber2004numerical,museth2017novel}.
The presented method is more specific in the treating of the sign condition but follows these methods in essence.

The proposed method does not achieve the order of accuracy expected from the Godunov scheme and WENO derivatives.
The degenerate order of accuracy arises from the narrowband being initialized with an exact signed distance transform, which is only first-order accurate.
It was assumed that multiple iterations of the fast sweeping method would improve this order.
While multiple iterations did improve the order of accuracy (Figure~\ref{sdt:fig:noise:proposed}), this did not lead to the same accuracy as the WENO derivatives.
Most likely this is caused by a contradiction in the upwind scheme where interior narrowband voxels are solved upwind of exterior narrowband voxels and visa versa, causing the causality condition to be invalidated.
A high order initialization method is needed for the problem.

The major issue for high-order signed distance transforms of sampled signals is that the problem does not permit a unique solution.
One attempt to resolve this non-uniqueness would be to add a constraint on the higher derivatives, such as minimizing the Laplacian (curvature) of the embedding while enforcing the Eikonal, recovery, and high order conditions.
This could be solved in the narrowband and extended outwards using FMM, FSM, or FIM.
Alternatively, the surface could be embedded with a function different from the signed distance transform~\cite{whitaker2000reducing,lempitsky2010surface}.
Distances in the narrowband could be solved using closest point methods~\cite{macdonald2008level} and extended using FFM, FSM, or FIM.
Either way, modifying the problem to permit a unique solution is the central challenge.

\section{Conclusion}
\label{sdt:sec:conclusion}
The high-order signed distance transform of a sampled signal is a transform satisfying the Eikonal equation, recovery condition, and has an order of accuracy greater than unity.
This transform is better than the traditional ``exact'' signed distance transform of binary sampled signals because of the increased order of accuracy (in the $\ell^1$ norm).
A method is proposed for the transform based on the fast sweeping method.
However, arbitrary order of accuracy is not achieved with the proposed method because of the low quality initialization method.
An additional condition is needed to make the problem unique.


%






\bibliographystyle{IEEEtran}
\bibliography{IEEEabrv,sdt.bib}

\end{document}